\documentclass{article}

 \usepackage[main, final]{neurips_2026}
\usepackage{hyperref}
\usepackage{tikz}
\usepackage{xcolor}
\usepackage{amsmath}

\usepackage{booktabs}
\usepackage{tabularx}
\usepackage{array}

\usetikzlibrary{
    arrows.meta,
    positioning,
    calc,
    fit,
    backgrounds
}

\definecolor{graphblue}{RGB}{40,96,160}
\definecolor{looporange}{RGB}{218,124,48}
\definecolor{verifygreen}{RGB}{55,130,90}
\definecolor{lightblue}{RGB}{231,240,251}
\definecolor{lightgreen}{RGB}{230,244,235}
\definecolor{lightgray}{RGB}{245,245,245}

\tikzset{
    >={Latex[length=2.5mm,width=1.8mm]},
    request/.style={
        rectangle,
        rounded corners=3pt,
        draw=black,
        very thick,
        fill=lightgray,
        align=center,
        minimum width=2.8cm,
        minimum height=0.9cm,
        font=\bfseries
    },
    graphcontrol/.style={
        rectangle,
        rounded corners=3pt,
        draw=graphblue,
        very thick,
        fill=lightblue,
        align=center,
        minimum width=4.2cm,
        minimum height=1cm,
        font=\bfseries
    },
    process/.style={
        rectangle,
        rounded corners=2pt,
        draw=graphblue,
        thick,
        fill=white,
        align=center,
        minimum width=1.9cm,
        minimum height=0.85cm,
        font=\small
    },
    verification/.style={
        rectangle,
        rounded corners=2pt,
        draw=verifygreen,
        very thick,
        fill=lightgreen,
        align=center,
        minimum width=1.9cm,
        minimum height=0.85cm,
        font=\small\bfseries
    },
    mainarrow/.style={
        ->,
        very thick,
        draw=graphblue
    },
    looparrow/.style={
        ->,
        very thick,
        draw=looporange,
        rounded corners=5pt
    },
    looplabel/.style={
        fill=white,
        inner sep=2pt,
        text=looporange,
        font=\scriptsize\bfseries,
        align=center
    }
}

\usepackage{amsmath}
\usepackage{amssymb}

\title{Forward-Deployed Full-Stack Engineering for Autonomous Cloud MLOps}

\author{%
  Sagar Srinivas Sakhinana, Venkataramana Runkana \\
  \texttt{sagar.sakhinana@tcs.com, venkat.runkana@tcs.com} \\
  Tata Research Development and Design Centre
}

\begin{document}

\maketitle

%%%%%%%%%%%%%%%%%%%%%%%%%%%%%%%%%%%%%%%%%%%%%
\begin{abstract}
Across industries, machine-learning systems support applications ranging from prediction and anomaly detection to forecasting, optimization, and scheduling, yet operationalizing these systems requires coordinating application development, model pipelines, cloud infrastructure, security, deployment, monitoring, retraining, recovery, and rollback. We present an evidence-gated multi-agent framework for transforming a natural-language MLOps cloud engineering task into a verified repository and operational cloud deployment. The framework combines \emph{graph engineering}, \emph{loop engineering}, and \emph{agent harness engineering}. A stateful Graph Orchestrator coordinates specialized agents for repository generation, review, execution, verification, release, and monitoring while governing workflow dependencies, evidence gates, retry bounds, recovery paths, and termination. Consequential lifecycle transitions proceed only when their required predicates are supported by verifiable execution or runtime evidence. Verification failures activate bounded reflection, repair, and re-verification, while runtime evidence of failure, drift, degradation, or policy violation can trigger bounded adaptation, recovery, or rollback. Agent harness engineering constrains repository generation, review, and repair, artifact execution, and cloud operations through controlled capabilities and isolated execution environments. We realize the framework on Google Cloud Platform and evaluate repository completeness, controlled execution, evidence-gated transitions, cloud promotion, and bounded recovery. Our experimental results show that the framework prevents unsupported lifecycle transitions and drives each run toward either a verified operational deployment or an auditable terminal failure.
\end{abstract}
%%%%%%%%%%%%%%%%%%%%%%%%%%%%%%%%%%%%%%%%%%%%%

%%%%%%%%%%%%%%%%%%%%%%%%%%%%%%%%%%%%%%%%%%%%%
\section{Introduction}
\label{sec:introduction}
Industries ranging from oil and gas, pharmaceuticals, semiconductor manufacturing, energy, automotive, FMCG, aerospace, mining, metallurgy and so on increasingly depend on applications such as predictive maintenance, anomaly and defect detection, demand forecasting, quality prediction, process optimization, inventory planning, transport and distribution routing, scheduling  and many others. Despite their application diversity, they share a common operational lifecycle spanning reproducible pipelines, cloud infrastructure provisioning, testing, security scanning, deployment, observability, retraining, failure recovery, auditability, and provenance(data lineage, model versions, and execution logs). This raises a broader question: how much of this lifecycle can be generated and operated by AI agents without trusting their claims of success? We argue that agent autonomy should instead depend on independently verifiable evidence from execution traces, deterministic compliance checks, deployment health, and runtime telemetry. Recent work has consequently shifted from isolated prompts toward the holistic engineering of agentic systems through structured execution environments, computation graphs, persistent context, bounded feedback loops, and trajectory-grounded evaluation~\citep{hassan2025agenticse,yue2026workflowgraphs,macedo2026promptgraph,macedo2026loops,zhang2025ace,anthropic2025context,anthropic2025harness,anthropic2026evals}. Nevertheless, prior work has largely focused on reasoning, code synthesis, repository modification, graph/workflow optimization, or bounded task completion~\citep{ding2026nl2repo, zhang2026repozero, zhou2026agents4d, wang2026tracestrust, rabanser2026reliability, wang2026reflect} rather than the end-to-end transformation of a natural-language long-horizon cloud engineering task - for example, ``Build, deploy, and operate a time series anomaly-detection pipeline on cloud infrastructure using Kubernetes, with continuous monitoring, drift detection, automated retraining, autoscaling, and rollback"—into a verified operational deployment. An end-to-end MLOps system spans continuous integration, continuous delivery, and continuous training (CI/CD/CT)~\citep{baylor2017tfx,kreuzberger2023mlops}. CI covers data acquisition, validation, feature engineering, and evaluation; CT covers training, hyperparameter tuning, experiment tracking, model registration, drift detection, and retraining; and CD covers serving, monitoring, and retirement~\citep{sculley2015technicaldebt,baylor2017tfx,kreuzberger2023mlops}. Cloud operationalization further requires infrastructure provisioning, identity and access management, secrets management, immutable artifact versioning, network isolation, Policy as Code (PaC) rollout, elastic scaling, observability, recovery, and rollback. We study autonomous cloud MLOps engineering through a representative cloud application comprising frontend and backend services. The frontend provides a secure web interface for user interaction and result visualization, while the backend exposes authenticated application and inference APIs and manages authorization, secrets, administrative operations, and audit retrieval. Shared telemetry, provenance, and audit records provide verifiable evidence across both tiers. AI Agents must build and test artifacts, scan source code and container images, publish immutable versions, provision infrastructure, reconcile deployment state, evaluate models, and verify releases before promotion. The lifecycle is cyclic and non-linear: production telemetry may trigger retraining, re-evaluation, staged rollout, failover, rollback, or retirement. Because these stages are interdependent, success at any individual stage does not establish that the resulting cloud system is secure, resilient, or operational. In agent-driven, long-horizon workflows, apparent completion may conceal stale state, failed handoffs, or policy violations that emerge only across multi-step execution and verification of the resulting cloud state~\citep{bousetouane2026hob}. This motivates our central research question:

\begin{quote}
\emph{Can a multi-agent system reliably transform a natural-language MLOps cloud engineering task into a verified, operational cloud deployment while requiring verifiable evidence for every consequential lifecycle transition?}
\end{quote}

To address this question, we present an agentic AI framework that transforms an end-user cloud engineering task into a verified operational cloud deployment. The framework generates a deployable repository comprising application code, data-processing and training pipelines, tests, container definitions, CI/CD workflows, infrastructure as code, Helm charts, Kubernetes manifests, GitOps configurations, security policies, observability, retraining, recovery, and rollback artifacts. Generated artifacts are inspected and executed in controlled sandbox environments using dependency, syntax, testing, security, policy, provenance, and runtime checks. Artifacts satisfying these predicates are built, scanned, signed, published, deployed to staging, and subsequently verified against live cloud infrastructure before promotion. Verification failures produce execution evidence that drives bounded correction and re-execution, while consequential lifecycle transitions proceed only when their required evidence predicates are satisfied. We operationalize evidence-gated autonomous MLOps through three complementary mechanisms: graph engineering, loop engineering, and agent harness engineering. Graph engineering represents the lifecycle as a stateful execution graph in which a Graph Orchestrator manages workflow state, dependencies, evidence gates, retry bounds, recovery paths, and termination. The graph coordinates specialized agents for repository generation, repository review, execution, verification, release, and monitoring, with production promotion conditioned on verified evidence from the preceding lifecycle stages. Loop engineering defines bounded control cycles for correction and runtime adaptation. When verification fails, the resulting evidence is routed to a Reflection Agent for diagnosis and a Repair Agent for artifact correction; corrected artifacts then re-enter repository review, execution, and verification before deployment proceeds. Runtime evidence of failure, drift, degradation, or policy violation similarly activates reflection and repair, after which the corrected system re-enters the appropriate verification, deployment, and runtime-validation stages. Retry bounds and terminal conditions are enforced by the Graph Orchestrator. Agent harness engineering constrains agent access to tools, execution environments, and operational systems. Repository generation, review, reflection, and repair agents operate through controlled VS Code and Chrome sandbox environments; artifact execution occurs

\newpage

%%%%%%%%%%%%%%%%%%%%%%%%%%%%%%%%%%%%%%%%%%%%%
\begin{figure*}[ht!]
\centering

\begin{tikzpicture}[
node distance=3.2mm,
stage/.style={
    graphcontrol,
    minimum width=7.4cm,
    text width=7.0cm,
    minimum height=0.70cm,
    inner ysep=3pt,
    font=\scriptsize,
    align=center
},
activity/.style={
    process,
    minimum width=7.0cm,
    text width=6.6cm,
    minimum height=0.70cm,
    inner ysep=3pt,
    font=\scriptsize,
    align=center
},
gate/.style={
    verification,
    minimum width=7.4cm,
    text width=7.0cm,
    minimum height=0.70cm,
    inner ysep=3pt,
    font=\scriptsize,
    align=center
}
]

% --------------------------------------------------------------
% Natural-language requirement
% --------------------------------------------------------------
\node[
    request,
    minimum width=5.8cm,
    minimum height=0.70cm,
    inner ysep=3pt,
    font=\scriptsize,
    align=center
] (request) {
    \textbf{Natural-Language MLOps Cloud Engineering Task}
};

% --------------------------------------------------------------
% Multi-agent engineering
% --------------------------------------------------------------
\node[
    stage,
    below=of request,
    minimum height=1.25cm
] (engineering) {
    \textbf{Multi-Agent Engineering}\\[-1pt]
    \textbf{Graph:} long-horizon lifecycle control
    \quad $\vert$ \quad
    \textbf{Loop:} bounded correction and adaptation\\[-1pt]
    \textbf{Harness:} controlled capabilities and execution
};

% --------------------------------------------------------------
% Generated repository and deployment artifacts
% --------------------------------------------------------------
\node[
    stage,
    below=of engineering
] (artifacts) {
    \textbf{Generated Repository + Deployment Artifacts}
};

% --------------------------------------------------------------
% Controlled artifact execution
% --------------------------------------------------------------
\node[
    activity,
    below=of artifacts
] (execution) {
    \textbf{Controlled Artifact Execution}
};

% --------------------------------------------------------------
% Artifact and release evidence
% --------------------------------------------------------------
\node[
    activity,
    below=of execution,
    minimum height=0.85cm
] (releaseevidence) {
    \textbf{Artifact + Release Evidence}\\[-1pt]
    {\scriptsize tests, scans, policy checks, provenance, and execution traces}
};

% --------------------------------------------------------------
% Artifact and release verification
% --------------------------------------------------------------
\node[
    gate,
    below=of releaseevidence,
    minimum height=0.85cm
] (releaseverification) {
    \textbf{Artifact + Release Verification}\\[-1pt]
    {\scriptsize required predicates gate deployment and promotion}
};

% --------------------------------------------------------------
% Operational cloud deployment
% --------------------------------------------------------------
\node[
    activity,
    below=of releaseverification
] (deployment) {
    \textbf{Operational Cloud Deployment}
};

% --------------------------------------------------------------
% Runtime evidence
% --------------------------------------------------------------
\node[
    activity,
    below=of deployment,
    minimum height=0.85cm
] (runtimeevidence) {
    \textbf{Runtime Evidence}\\[-1pt]
    {\scriptsize deployment health, runtime telemetry, drift, degradation, and policy signals}
};

% --------------------------------------------------------------
% Runtime verification
% --------------------------------------------------------------
\node[
    gate,
    below=of runtimeevidence,
    minimum height=0.85cm
] (runtimeverification) {
    \textbf{Runtime Verification}\\[-1pt]
    {\scriptsize operational predicates gate continued operation and completion}
};

% --------------------------------------------------------------
% Verified operational cloud deployment
% --------------------------------------------------------------
\node[
    verification,
    below=of runtimeverification,
    minimum width=5.8cm,
    minimum height=0.70cm,
    inner ysep=3pt,
    font=\scriptsize,
    align=center
] (completion) {
    \textbf{Verified Operational Cloud Deployment}
};

% --------------------------------------------------------------
% Forward lifecycle
% --------------------------------------------------------------
\draw[mainarrow] (request) -- (engineering);
\draw[mainarrow] (engineering) -- (artifacts);
\draw[mainarrow] (artifacts) -- (execution);
\draw[mainarrow] (execution) -- (releaseevidence);
\draw[mainarrow] (releaseevidence) -- (releaseverification);

\draw[mainarrow]
(releaseverification)
-- node[
    right,
    font=\scriptsize,
    text=verifygreen
] {satisfied}
(deployment);

\draw[mainarrow] (deployment) -- (runtimeevidence);
\draw[mainarrow] (runtimeevidence) -- (runtimeverification);

\draw[mainarrow]
(runtimeverification)
-- node[
    right,
    font=\scriptsize,
    text=verifygreen
] {satisfied}
(completion);

% --------------------------------------------------------------
% Bounded correction loop
% Artifact/release verification failure -> multi-agent engineering
% --------------------------------------------------------------
\coordinate (correctlane)
at ($(releaseverification.east)+(5mm,0)$);

\draw[looparrow]
(releaseverification.east)
-- (correctlane)
|- node[
    right,
    pos=0.25,
    font=\scriptsize,
    text=looporange
] {correct}
(engineering.east);

% --------------------------------------------------------------
% Bounded runtime adaptation loop
% Runtime verification trigger -> multi-agent engineering
% --------------------------------------------------------------
\coordinate (adaptlane)
at ($(runtimeverification.west)+(-5mm,0)$);

\draw[looparrow]
(runtimeverification.west)
-- (adaptlane)
|- node[
    left,
    pos=0.20,
    font=\scriptsize,
    text=looporange
] {adapt}
(engineering.west);

\end{tikzpicture}

\caption{The figure shows how the evidence-gated autonomous cloud MLOps transforms a natural-language engineering task into a verified operational cloud deployment. Graph engineering governs
long-horizon lifecycle, loop engineering provides bounded artifact correction and runtime adaptation, agent harness engineering constrains execution, and verified evidence gates consequential lifecycle transitions.
}
\label{fig:agentic-mlops-workflow}
\vspace{-3mm}
\end{figure*}
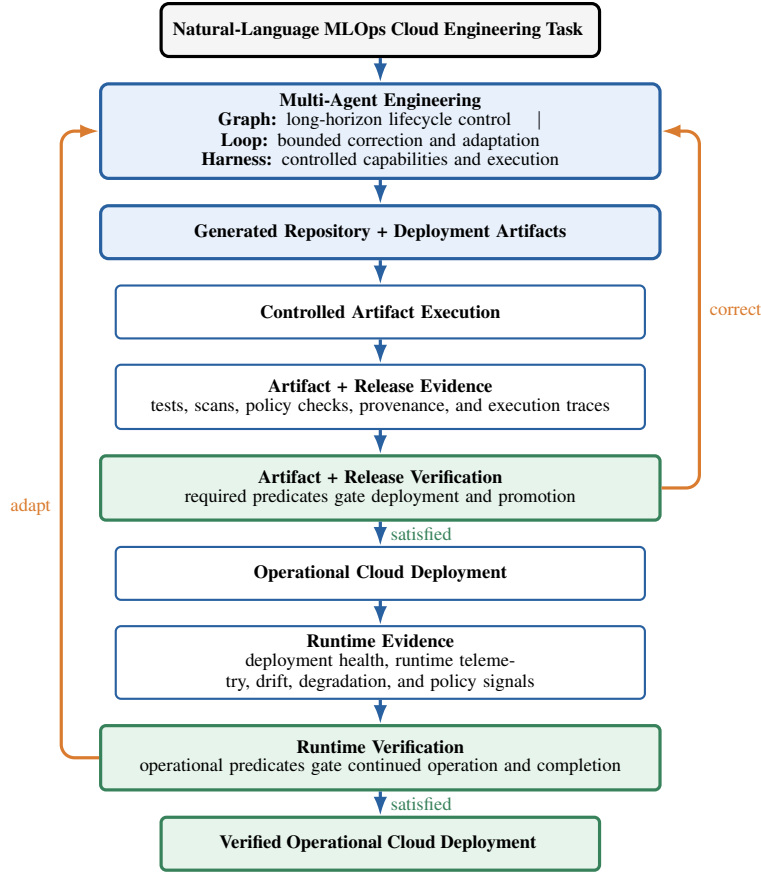
\begin{figure*}[t]
\centering

\begin{tikzpicture}[
node distance=6mm and 4mm,
agent/.style={
    process,
    minimum width=2.25cm,
    text width=1.95cm,
    minimum height=1.0cm,
    font=\scriptsize,
    align=center
},
verifyagent/.style={
    verification,
    minimum width=2.25cm,
    text width=1.95cm,
    minimum height=1.0cm,
    font=\scriptsize,
    align=center
},
environment/.style={
    process,
    fill=lightgray,
    minimum height=0.9cm,
    font=\scriptsize,
    align=center
}
]

% ------------------------------------------------------------------
% Natural-language requirement
% ------------------------------------------------------------------
\node[
    request,
    minimum width=4.8cm,
    font=\scriptsize\bfseries
] (requirement) {
    Natural-Language MLOps Requirement
};

% ------------------------------------------------------------------
% Graph orchestrator
% ------------------------------------------------------------------
\node[
    graphcontrol,
    below=4mm of requirement,
    minimum width=7.0cm,
    text width=6.55cm,
    minimum height=1.05cm,
    font=\scriptsize,
    align=center
] (orchestrator) {
    \textbf{Graph Orchestrator}
    \\[-0.2mm]
    State, dependencies, evidence gates, retry bounds,
    recovery paths, and termination
};

\draw[mainarrow] (requirement) -- (orchestrator);

% ------------------------------------------------------------------
% First lifecycle row
% Generation -> Review -> Execution -> Artifact Verification
%
% Compact vertical gap retained only for clear orchestrator entry.
% ------------------------------------------------------------------
\node[
    agent,
    below=8mm of orchestrator,
    xshift=-1.325cm
] (review) {
    \textbf{Repository\\Review Agent}
};

\node[
    agent,
    left=of review
] (generation) {
    \textbf{Repository\\Generation Agent}
};

\node[
    agent,
    right=of review
] (execution) {
    \textbf{Execution Agent}
};

\node[
    verifyagent,
    right=of execution
] (artifactverification) {
    \textbf{Verification Agent}
    \\[-0.2mm]
    Artifact checks
};

% ------------------------------------------------------------------
% Orchestrator entry into lifecycle
% ------------------------------------------------------------------
\coordinate (graphentry)
    at ($(generation.west)+(-5mm,0)$);

\draw[mainarrow]
    (orchestrator.west)
    -| (graphentry)
    -- (generation.west);

% ------------------------------------------------------------------
% First-row lifecycle flow
% ------------------------------------------------------------------
\draw[mainarrow] (generation) -- (review);
\draw[mainarrow] (review) -- (execution);
\draw[mainarrow] (execution) -- (artifactverification);

% ------------------------------------------------------------------
% Second lifecycle row
% Staging -> Live-cloud Verification -> Promotion -> Monitoring
%
% Reduced gap while retaining room for the verified transition.
% ------------------------------------------------------------------
\node[
    agent,
    below=6mm of artifactverification,
    minimum height=1.12cm
] (stagingrelease) {
    \textbf{Release Agent}
    \\[-0.2mm]
    Build, scan, sign,\\
    publish, and stage
};

\node[
    verifyagent,
    left=of stagingrelease,
    minimum height=1.12cm
] (liveverification) {
    \textbf{Verification Agent}
    \\[-0.2mm]
    Live-cloud\\
    verification
};

\node[
    agent,
    left=of liveverification,
    minimum height=1.12cm
] (promotion) {
    \textbf{Release Agent}
    \\[-0.2mm]
    Production\\
    promotion
};

\node[
    agent,
    left=of promotion,
    minimum height=1.12cm
] (monitoring) {
    \textbf{Monitoring Agent}
    \\[-0.2mm]
    Runtime evidence
};

% ------------------------------------------------------------------
% Evidence-gated release lifecycle
% ------------------------------------------------------------------
\draw[mainarrow]
    (artifactverification)
    -- node[
        right,
        font=\scriptsize,
        text=verifygreen
    ] {verified}
    (stagingrelease);

\draw[mainarrow]
    (stagingrelease)
    -- (liveverification);

\draw[mainarrow]
    (liveverification)
    -- node[
        above,
        font=\scriptsize,
        text=verifygreen
    ] {verified}
    (promotion);

\draw[mainarrow]
    (promotion)
    -- (monitoring);

% ------------------------------------------------------------------
% Reflection and repair agents
%
% Enough clearance is retained for live-cloud verification failure
% and for the Loop Engineering boundary title.
% ------------------------------------------------------------------
\node[
    agent,
    below=9mm of liveverification
] (reflection) {
    \textbf{Reflection Agent}
    \\[-0.3mm]
    Diagnose failure
};

\node[
    agent,
    left=of reflection
] (repair) {
    \textbf{Repair Agent}
    \\[-0.3mm]
    Correct artifacts
};

% ------------------------------------------------------------------
% Artifact-verification failure -> Reflection
% ------------------------------------------------------------------
\coordinate (artifactfaillane)
    at ($(stagingrelease.east)+(6mm,0)$);

\coordinate (artifactfailturn)
    at (artifactfaillane |- reflection.east);

\draw[looparrow]
    (artifactverification.east)
    -- (artifactfaillane |- artifactverification.east)
    -- node[
        looplabel,
        right,
        pos=0.52
    ] {predicate failed}
    (artifactfailturn)
    -- (reflection.east);

% ------------------------------------------------------------------
% Live-cloud verification failure -> Reflection
% ------------------------------------------------------------------
\draw[looparrow]
    (liveverification.south)
    -- node[
        looplabel,
        right
    ] {predicate failed}
    (reflection.north);

% ------------------------------------------------------------------
% Runtime-triggered adaptation
%
% This 6 mm clearance is retained intentionally. It prevents the
% runtime feedback lane from colliding with Reflection/Repair boxes
% or the Agent Harness Engineering boundary below.
% ------------------------------------------------------------------
\coordinate (runtimebase)
    at ($(reflection.south)+(0,-6mm)$);

\coordinate (monitoringbase)
    at (monitoring.south |- runtimebase);

\coordinate (reflectionbase)
    at (reflection.south |- runtimebase);

\draw[looparrow]
    (monitoring.south)
    -- node[
        looplabel,
        left,
        pos=0.55
    ] {runtime trigger}
    (monitoringbase)
    -- (reflectionbase)
    -- (reflection.south);

% ------------------------------------------------------------------
% Reflection -> Repair
% ------------------------------------------------------------------
\draw[looparrow]
    (reflection.west)
    -- node[
        looplabel,
        above
    ] {diagnosis}
    (repair.east);

% ------------------------------------------------------------------
% Repair -> Graph-controlled re-entry
% ------------------------------------------------------------------
\coordinate (reentrylane)
    at ($(generation.west)+(-10mm,0)$);

\coordinate (repairreentry)
    at (reentrylane |- repair.west);

\draw[looparrow]
    (repair.west)
    -- node[
        looplabel,
        below
    ] {correct}
    (repairreentry)
    |- node[
        looplabel,
        left,
        pos=0.28
    ] {re-enter graph}
    (orchestrator.west);

% ------------------------------------------------------------------
% Agent Harness Engineering
%
% Reduced from 19 mm to 17 mm. This is close enough to avoid wasted
% vertical space while still clearing the runtime feedback lane and
% the Harness boundary title.
% ------------------------------------------------------------------
\node[
    environment,
    below=17mm of $(repair.south)!0.5!(reflection.south)$,
    minimum width=3.7cm,
    text width=3.35cm
] (gvisor) {
    \textbf{GKE Sandbox\\with gVisor}
    \\[0.2mm]
    Execution
};

\node[
    environment,
    left=6mm of gvisor,
    minimum width=4.3cm,
    text width=3.95cm
] (sharedtools) {
    \textbf{VS Code and Chrome Sandboxes}
    \\[0.2mm]
    Repository generation, review,
    reflection, and repair
};

\node[
    environment,
    right=6mm of gvisor,
    minimum width=4.6cm,
    text width=4.20cm
] (cloud) {
    \textbf{Controlled Cloud Infrastructure}
    \\[0.2mm]
    Live-cloud verification, release,
    monitoring, recovery, and rollback
};

% ------------------------------------------------------------------
% Visual boundaries
% ------------------------------------------------------------------
\begin{scope}[on background layer]

% ------------------------------------------------------------------
% Graph Engineering boundary
% ------------------------------------------------------------------
\node[
    draw=graphblue,
    dashed,
    rounded corners=5pt,
    inner sep=4mm,
    fit={
        (orchestrator)
        (generation)
        (review)
        (execution)
        (artifactverification)
        (stagingrelease)
        (liveverification)
        (promotion)
        (monitoring)
        (reflection)
        (repair)
    }
] (graphboundary) {};

% ------------------------------------------------------------------
% Loop Engineering boundary
% ------------------------------------------------------------------
\node[
    draw=looporange,
    dashed,
    rounded corners=4pt,
    inner sep=3mm,
    fit=(reflection)(repair)
] (loopboundary) {};

% ------------------------------------------------------------------
% Agent Harness Engineering boundary
% ------------------------------------------------------------------
\node[
    draw=gray,
    dashed,
    rounded corners=5pt,
    inner sep=3.5mm,
    fit=(sharedtools)(gvisor)(cloud)
] (harnessboundary) {};

\end{scope}

% ------------------------------------------------------------------
% Boundary labels
% ------------------------------------------------------------------

% Graph Engineering
\node[
    anchor=west,
    fill=white,
    inner xsep=2pt,
    inner ysep=1pt,
    text=graphblue,
    font=\scriptsize\bfseries
] at ($(graphboundary.north west)+(3mm,0)$) {
    Graph Engineering
};

% Loop Engineering
\node[
    anchor=west,
    fill=white,
    inner xsep=2pt,
    inner ysep=1pt,
    text=looporange,
    font=\scriptsize\bfseries
] at ($(loopboundary.north west)+(3mm,0)$) {
    Loop Engineering
};

% Agent Harness Engineering
\node[
    anchor=west,
    fill=white,
    inner xsep=2pt,
    inner ysep=1pt,
    text=black,
    font=\scriptsize\bfseries
] at ($(harnessboundary.north west)+(3mm,0)$) {
    Agent Harness Engineering
};

\end{tikzpicture}
\caption{The figure shows the multi-agent architecture for evidence-gated autonomous cloud MLOps.
The Graph Orchestrator manages workflow state, dependencies, evidence gates,
retry bounds, recovery paths, and termination across specialized agents.
Verification failures and adverse runtime evidence activate bounded reflection
and repair. Agent Harness Engineering constrains agent access through controlled
sandboxes and cloud infrastructure.}
\label{fig:multi-agent-mlops-architecture}
\vspace{-5mm}
\end{figure*}
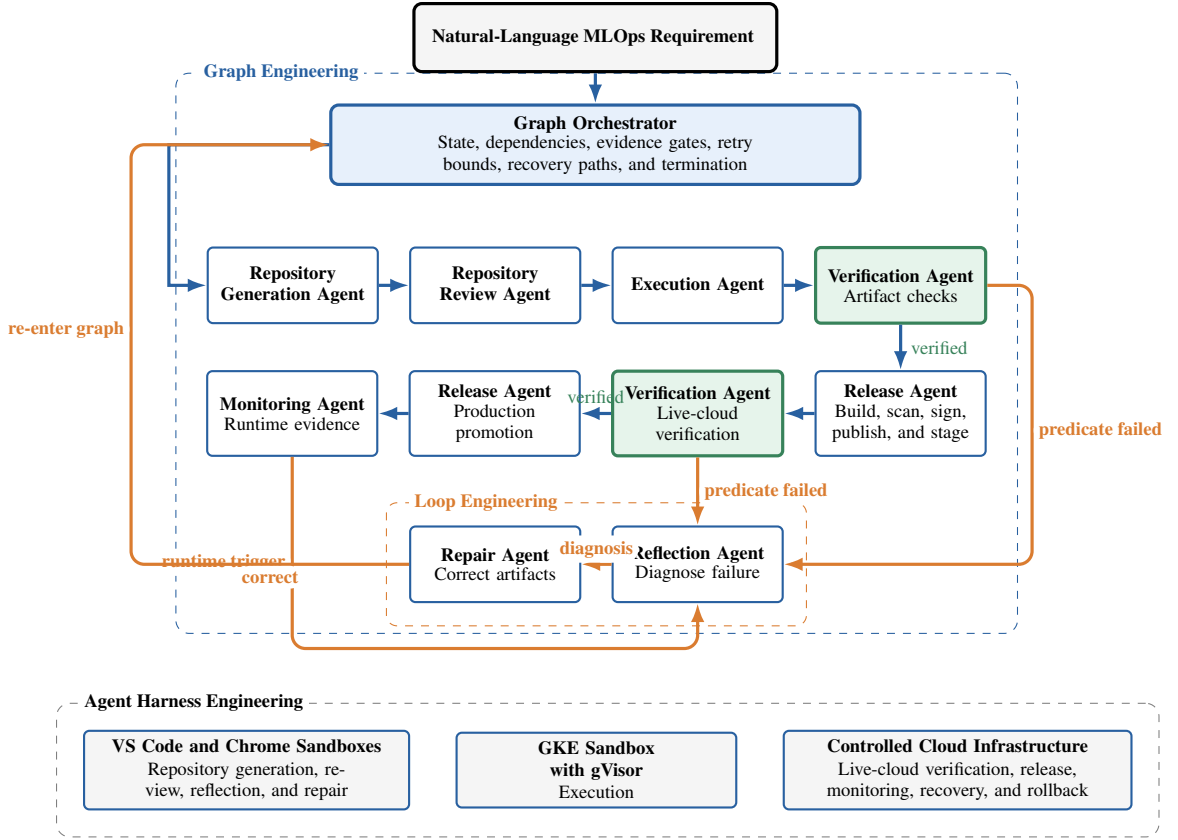

%%%%%%%%%%%%%%%%%%%%%%%%%%%%%%%%%%%%%%%%%%%%%

within a Google Kubernetes Engine(GKE) Sandbox with gVisor; and release and monitoring agents interact with controlled cloud infrastructure supporting staging, production, recovery, and rollback. Graph engineering~\citep{zhuge2024languageagents,zhang2024aflow} structures the end-to-end cloud MLOps workflow and its transitions, loop engineering~\citep{shinn2023reflexion,gou2023critic} provides bounded correction and adaptation, and agent harness engineering~\citep{zhong2026aiharness,lin2026agenticharness} defines the controlled operational context in which agents act. Figure~\ref{fig:agentic-mlops-workflow} presents the evidence-gated lifecycle from a natural-language MLOps cloud engineering task through artifact generation and execution, release, deployment, runtime verification, and verified operation, while Figure~\ref{fig:multi-agent-mlops-architecture} presents its multi-agent realization through graph-controlled execution, bounded correction and adaptation, and controlled execution environments. Our work makes four contributions:

\begin{itemize}
\item We formulate autonomous cloud MLOps as transforming a natural-language task into a verified repository and operational deployment through evidence-gated lifecycle transitions.
\item We introduce a stateful execution graph governed by a Graph Orchestrator that coordinates agents for repository generation, review, execution, verification, release, and monitoring while managing state, dependencies, evidence gates, retry bounds, recovery paths, and termination.
\item We introduce bounded loop engineering for evidence-driven correction and runtime adaptation. Failed verification invokes reflection, diagnosis, repair, and re-verification, while runtime failures, drift, degradation, or policy violations trigger bounded recovery and adaptation.
\item We realize agent harness engineering on Google Cloud Platform(GCP) through controlled execution environments: VS Code and Chrome sandboxes support repository generation, review, reflection, and repair; a gVisor-isolated GKE Sandbox supports artifact execution; and cloud infrastructure supports live-cloud verification, release, monitoring, recovery, and rollback. The generated repository includes application and pipeline code, tests, CI/CD, infrastructure as code, container and Kubernetes artifacts, GitOps configurations, security policies, provenance, observability, retraining, recovery, and rollback.
\end{itemize}

Our evaluation covers repository completeness, isolated execution, evidence-gated transitions, cloud promotion, and bounded recovery. We assess whether each run reaches a verified deployment or an auditable terminal failure. Our experimental results support the proposed framework and its underlying hypotheses, which we evaluate and discuss in the subsequent sections.
%%%%%%%%%%%%%%%%%%%%%%%%%%%%%%%%%%%%%%%%%%%%

%%%%%%%%%%%%%%%%%%%%%%%%%%%%%%%%%%%%%%%%%%%%%
%%%%%%%%%%%%%%%%%%%%%%%%%%%%%%%%%%%%%%%%%%%%%

%%%%%%%%%%%%%%%%%%%%%%%%%%%%%%%%%%%%%%%%%%%%%
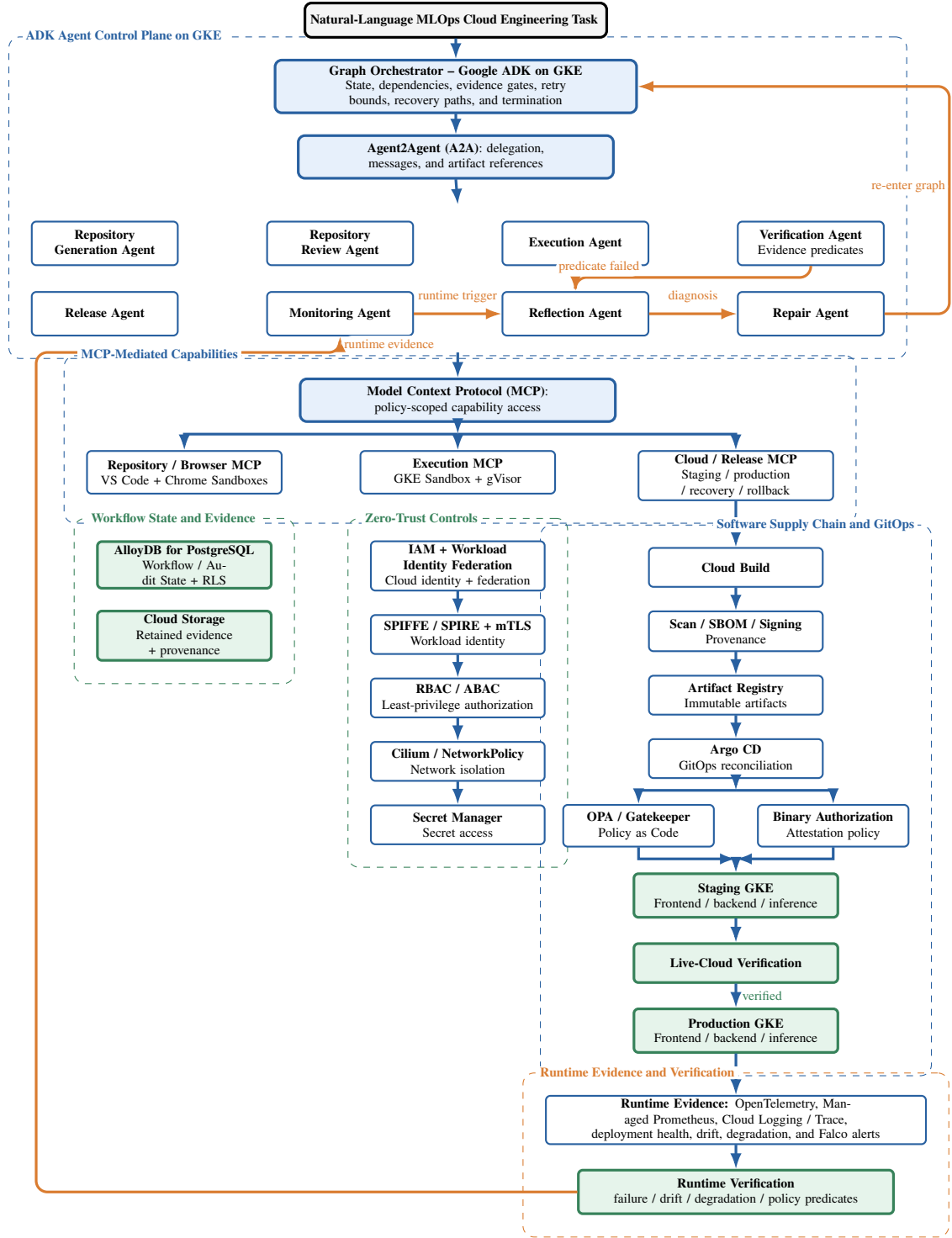
\begin{figure*}[ht!]
\centering

\begin{tikzpicture}[
    scale=0.729,
    transform shape,
    looplabel/.append style={font=\footnotesize},
    agent/.style={
        process,
        minimum width=3.15cm,
        text width=2.70cm,
        minimum height=0.94cm,
        font=\footnotesize,
        inner sep=2pt,
        align=center
    },
    protocol/.style={
        graphcontrol,
        minimum width=6.8cm,
        text width=6.35cm,
        minimum height=0.92cm,
        font=\footnotesize,
        inner sep=2pt,
        align=center
    },
    capability/.style={
        process,
        minimum width=4.20cm,
        text width=3.75cm,
        minimum height=0.92cm,
        font=\footnotesize,
        inner sep=2pt,
        align=center
    },
    plane/.style={
        process,
        minimum width=3.75cm,
        text width=3.30cm,
        minimum height=0.84cm,
        font=\footnotesize,
        inner sep=2pt,
        align=center
    },
    store/.style={
        verification,
        minimum width=3.75cm,
        text width=3.30cm,
        minimum height=0.84cm,
        font=\footnotesize,
        inner sep=2pt,
        align=center
    }
]

% ==================================================================
% Natural-language task
% ==================================================================
\node[
    request,
    minimum width=5.8cm,
    minimum height=0.74cm,
    font=\footnotesize\bfseries,
    align=center
] (req) {
    Natural-Language MLOps Cloud Engineering Task
};

% ==================================================================
% Graph orchestration
% ==================================================================
\node[
    graphcontrol,
    below=4.5mm of req,
    minimum width=7.8cm,
    text width=7.35cm,
    minimum height=1.12cm,
    font=\footnotesize,
    align=center
] (orch) {
    \textbf{Graph Orchestrator -- Google ADK on GKE}\\[-0.2mm]
    State, dependencies, evidence gates, retry bounds,
    recovery paths, and termination
};

\draw[mainarrow] (req) -- (orch);

% ==================================================================
% A2A
% ==================================================================
\node[
    protocol,
    below=4.5mm of orch
] (a2a) {
    \textbf{Agent2Agent (A2A)}:
    delegation, messages, and artifact references
};

\draw[mainarrow] (orch) -- (a2a);

% ==================================================================
% Specialized agents
% ==================================================================
\node[
    agent,
    below=9mm of a2a,
    xshift=-7.575cm
] (gen) {
    \textbf{Repository\\Generation Agent}
};

\node[
    agent,
    right=19mm of gen
] (review) {
    \textbf{Repository\\Review Agent}
};

\node[
    agent,
    right=19mm of review
] (exec) {
    \textbf{Execution Agent}
};

\node[
    agent,
    right=19mm of exec
] (verify) {
    \textbf{Verification Agent}\\[-0.2mm]
    Evidence predicates
};

\node[
    agent,
    below=5.5mm of gen
] (release) {
    \textbf{Release Agent}
};

\node[
    agent,
    below=5.5mm of review
] (monitor) {
    \textbf{Monitoring Agent}
};

\node[
    agent,
    below=5.5mm of exec
] (reflect) {
    \textbf{Reflection Agent}
};

\node[
    agent,
    below=5.5mm of verify
] (repair) {
    \textbf{Repair Agent}
};

% A2A enters the agent region
\coordinate (agenttop)
    at ($(review.north)!0.5!(exec.north)+(0,3.5mm)$);

\draw[mainarrow]
    (a2a.south) -- (agenttop);

% Verification failure -> reflection
\coordinate (verifyfail)
    at ($(verify.south)+(0,-2.5mm)$);

\draw[looparrow]
    (verify.south)
    -- (verifyfail)
    -| node[
        looplabel,
        above,
        pos=0.45
    ] {predicate failed}
    (reflect.north);

% Runtime trigger -> reflection
\draw[looparrow]
    (monitor.east)
    -- node[
        looplabel,
        above,
        yshift=1.5mm
    ] {runtime trigger}
    (reflect.west);

% Reflection -> repair
\draw[looparrow]
    (reflect.east)
    -- node[
        looplabel,
        above,
        yshift=1.5mm
    ] {diagnosis}
    (repair.west);

% Repair -> graph-controlled re-entry
\coordinate (reentrylane)
    at ($(repair.east)+(14mm,0)$);

\draw[looparrow]
    (repair.east)
    -- (reentrylane)
    |- node[
        looplabel,
        left,
        pos=0.28
    ] {re-enter graph}
    (orch.east);

% ==================================================================
% MCP capability layer
% ==================================================================
\node[
    protocol,
    below=9mm of $(monitor.south)!0.5!(reflect.south)$
] (mcp) {
    \textbf{Model Context Protocol (MCP)}:
    policy-scoped capability access
};

\coordinate (agentbot)
    at ($(monitor.south)!0.5!(reflect.south)+(0,-3.5mm)$);

\draw[mainarrow]
    (agentbot) -- (mcp.north);

% ==================================================================
% MCP-mediated controlled capabilities
% ==================================================================
\node[
    capability,
    below=6mm of mcp,
    xshift=-5.9cm
] (repoenv) {
    \textbf{Repository / Browser MCP}\\
    VS Code + Chrome Sandboxes
};

\node[
    capability,
    below=6mm of mcp
] (sandboxenv) {
    \textbf{Execution MCP}\\
    GKE Sandbox + gVisor
};

% IMPORTANT:
% This node defines the common centerline for the complete
% release/deployment/runtime path below.
\node[
    capability,
    below=6mm of mcp,
    xshift=6.0cm
] (cloudenv) {
    \textbf{Cloud / Release MCP}\\
    Staging / production / recovery / rollback
};

\coordinate (mcpbus)
    at ($(mcp.south)+(0,-2.5mm)$);

\draw[mainarrow]
    (mcp.south) -- (mcpbus);

\draw[mainarrow]
    (mcpbus) -| (repoenv.north);

\draw[mainarrow]
    (mcpbus) -| (sandboxenv.north);

\draw[mainarrow]
    (mcpbus) -| (cloudenv.north);

% ==================================================================
% Persistent workflow state and retained evidence
% ==================================================================
\node[
    store,
    below=10mm of repoenv
] (state) {
    \textbf{AlloyDB for PostgreSQL}\\
    Workflow / Audit State + RLS
};

\node[
    store,
    below=4.5mm of state
] (evidencestore) {
    \textbf{Cloud Storage}\\
    Retained evidence + provenance
};

% ==================================================================
% Zero-trust controls
% ==================================================================
\node[
    plane,
    below=10mm of sandboxenv
] (iam) {
    \textbf{IAM + Workload Identity Federation}\\
    Cloud identity + federation
};

\node[
    plane,
    below=5mm of iam
] (spiffe) {
    \textbf{SPIFFE / SPIRE + mTLS}\\
    Workload identity
};

\node[
    plane,
    below=5mm of spiffe
] (authz) {
    \textbf{RBAC / ABAC}\\
    Least-privilege authorization
};

\node[
    plane,
    below=5mm of authz
] (cilium) {
    \textbf{Cilium / NetworkPolicy}\\
    Network isolation
};

\node[
    plane,
    below=5mm of cilium
] (secrets) {
    \textbf{Secret Manager}\\
    Secret access
};

\draw[mainarrow] (iam) -- (spiffe);
\draw[mainarrow] (spiffe) -- (authz);
\draw[mainarrow] (authz) -- (cilium);
\draw[mainarrow] (cilium) -- (secrets);

% ==================================================================
% Software supply chain and GitOps
%
% Every node in this operational path is centered directly below
% Cloud / Release MCP. No xshift is used here.
% ==================================================================
\node[
    plane,
    below=10mm of cloudenv
] (build) {
    \textbf{Cloud Build}
};

\node[
    plane,
    below=5mm of build,
    minimum height=0.86cm
] (supply) {
    \textbf{Scan / SBOM / Signing}\\
    Provenance
};

\node[
    plane,
    below=5mm of supply
] (registry) {
    \textbf{Artifact Registry}\\
    Immutable artifacts
};

\node[
    plane,
    below=5mm of registry
] (argocd) {
    \textbf{Argo CD}\\
    GitOps reconciliation
};

% ==================================================================
% Deployment-policy gates
%
% Both policy gates branch symmetrically around the SAME operational
% centerline and rejoin on that centerline.
% ==================================================================
\node[
    plane,
    below=5.5mm of argocd,
    xshift=-2.10cm,
    minimum width=3.25cm,
    text width=2.83cm,
    minimum height=0.90cm
] (gate) {
    \textbf{OPA / Gatekeeper}\\
    Policy as Code
};

\node[
    plane,
    below=5.5mm of argocd,
    xshift=2.10cm,
    minimum width=3.25cm,
    text width=2.83cm,
    minimum height=0.90cm
] (binauth) {
    \textbf{Binary Authorization}\\
    Attestation policy
};

\node[
    verification,
    below=5.5mm of $(gate.south)!0.5!(binauth.south)$,
    minimum width=4.4cm,
    text width=3.95cm,
    minimum height=0.94cm,
    font=\footnotesize,
    align=center
] (staging) {
    \textbf{Staging GKE}\\[-0.2mm]
    Frontend / backend / inference
};

\node[
    verification,
    below=5.25mm of staging,
    minimum width=4.4cm,
    text width=3.95cm,
    minimum height=0.84cm,
    font=\footnotesize,
    align=center
] (liveverify) {
    \textbf{Live-Cloud Verification}
};

\node[
    verification,
    below=5.25mm of liveverify,
    minimum width=4.4cm,
    text width=3.95cm,
    minimum height=0.94cm,
    font=\footnotesize,
    align=center
] (production) {
    \textbf{Production GKE}\\[-0.2mm]
    Frontend / backend / inference
};

% ==================================================================
% Release pipeline
% ==================================================================
\draw[mainarrow] (cloudenv) -- (build);
\draw[mainarrow] (build) -- (supply);
\draw[mainarrow] (supply) -- (registry);
\draw[mainarrow] (registry) -- (argocd);

% Argo CD -> parallel deployment-policy gates
\coordinate (policybus)
    at ($(argocd.south)+(0,-2.5mm)$);

\draw[mainarrow]
    (argocd.south) -- (policybus);

\draw[mainarrow]
    (policybus) -| (gate.north);

\draw[mainarrow]
    (policybus) -| (binauth.north);

% Policy gates rejoin exactly at center
\coordinate (policyjoin)
    at ($(gate.south)!0.5!(binauth.south)+(0,-2.5mm)$);

\draw[mainarrow]
    (gate.south) |- (policyjoin);

\draw[mainarrow]
    (binauth.south) |- (policyjoin);

\draw[mainarrow]
    (policyjoin) -- (staging.north);

% ==================================================================
% Staging -> live verification -> production
% ==================================================================
\draw[mainarrow]
    (staging) -- (liveverify);

\draw[mainarrow]
    (liveverify)
    -- node[
        right,
        font=\footnotesize,
        text=verifygreen
    ] {verified}
    (production);

% ==================================================================
% Runtime observability, evidence, and verification
%
% No xshift here: both nodes remain on the exact same centerline
% as Cloud / Release MCP and the complete release path above.
% ==================================================================
\node[
    plane,
    below=9mm of production,
    minimum width=8.2cm,
    text width=7.65cm,
    minimum height=0.92cm
] (evidence) {
    \textbf{Runtime Evidence:}
    OpenTelemetry, Managed Prometheus, Cloud Logging / Trace,\\[-0.2mm]
    deployment health, drift, degradation, and Falco alerts
};

\node[
    verification,
    below=5.25mm of evidence,
    minimum width=6.8cm,
    text width=6.35cm,
    minimum height=0.94cm,
    font=\footnotesize,
    align=center
] (runtime) {
    \textbf{Runtime Verification}\\[-0.2mm]
    failure / drift / degradation / policy predicates
};

\draw[mainarrow]
    (production) -- (evidence);

\draw[mainarrow]
    (evidence) -- (runtime);

% ==================================================================
% Runtime evidence -> Monitoring Agent
% ==================================================================
\coordinate (runtimeleftlane)
    at ($(evidencestore.west)+(-13mm,0)$);

\coordinate (monitorentry)
    at ($(monitor.south)+(0,-3.5mm)$);

\draw[looparrow]
    (runtime.west)
    -| (runtimeleftlane)
    |- (monitorentry)
    -- node[
        looplabel,
        right,
        pos=0.55
    ] {runtime evidence}
    (monitor.south);

% ==================================================================
% Visual boundaries
% ==================================================================
\begin{scope}[on background layer]

% ADK/A2A agent control plane
\node[
    draw=graphblue,
    dashed,
    rounded corners=5pt,
    inner sep=3.5mm,
    fit={
        (orch)
        (a2a)
        (gen)
        (review)
        (exec)
        (verify)
        (release)
        (monitor)
        (reflect)
        (repair)
    }
] (agentboundary) {};

% MCP-mediated capability layer
\node[
    draw=graphblue,
    dashed,
    rounded corners=5pt,
    inner sep=3.5mm,
    fit={
        (mcp)
        (repoenv)
        (sandboxenv)
        (cloudenv)
    }
] (capabilityboundary) {};

% Persistent state and retained evidence
\node[
    draw=verifygreen,
    dashed,
    rounded corners=5pt,
    inner sep=3.5mm,
    fit={
        (state)
        (evidencestore)
    }
] (stateboundary) {};

% Zero-trust stack
\node[
    draw=verifygreen,
    dashed,
    rounded corners=5pt,
    inner sep=3.5mm,
    fit={
        (iam)
        (spiffe)
        (authz)
        (cilium)
        (secrets)
    }
] (zeroboundary) {};

% Software supply chain and GitOps
\node[
    draw=graphblue,
    dashed,
    rounded corners=5pt,
    inner sep=3.5mm,
    fit={
        (build)
        (supply)
        (registry)
        (argocd)
        (gate)
        (binauth)
        (staging)
        (liveverify)
        (production)
    }
] (supplyboundary) {};

% Runtime evidence and verification
\node[
    draw=looporange,
    dashed,
    rounded corners=5pt,
    inner sep=3.5mm,
    fit={
        (evidence)
        (runtime)
    }
] (runtimeboundary) {};

\end{scope}

% ==================================================================
% Boundary labels
% ==================================================================
\node[
    anchor=west,
    fill=white,
    inner xsep=2pt,
    inner ysep=1pt,
    text=graphblue,
    font=\footnotesize\bfseries
] at ($(agentboundary.north west)+(3mm,0)$) {
    ADK Agent Control Plane on GKE
};

\node[
    anchor=west,
    fill=white,
    inner xsep=2pt,
    inner ysep=1pt,
    text=graphblue,
    font=\footnotesize\bfseries
] at ($(capabilityboundary.north west)+(3mm,0)$) {
    MCP-Mediated Capabilities
};

\node[
    anchor=west,
    fill=white,
    inner xsep=2pt,
    inner ysep=1pt,
    text=verifygreen,
    font=\footnotesize\bfseries
] at ($(stateboundary.north west)+(3mm,0)$) {
    Workflow State and Evidence
};

\node[
    anchor=west,
    fill=white,
    inner xsep=2pt,
    inner ysep=1pt,
    text=verifygreen,
    font=\footnotesize\bfseries
] at ($(zeroboundary.north west)+(3mm,0)$) {
    Zero-Trust Controls
};

\node[
    anchor=east,
    fill=white,
    inner xsep=2pt,
    inner ysep=1pt,
    text=graphblue,
    font=\footnotesize\bfseries
] at ($(supplyboundary.north east)+(-3mm,0)$) {
    Software Supply Chain and GitOps
};

\node[
    anchor=west,
    fill=white,
    inner xsep=2pt,
    inner ysep=1pt,
    text=looporange,
    font=\footnotesize\bfseries
] at ($(runtimeboundary.north west)+(3mm,0)$) {
    Runtime Evidence and Verification
};

\end{tikzpicture}
\vspace{-6mm}
\caption{Google Cloud realization of the evidence-gated autonomous MLOps framework. The Graph Orchestrator, implemented with Google ADK on GKE, coordinates specialized agents through Agent2Agent (A2A), while the Model Context Protocol (MCP) provides policy-scoped access to repository and browser tools, isolated execution, and cloud deployment, release, recovery, and rollback operations.}
\label{fig:agentic-mlops-deployment-architecture}
\vspace{-2mm}
\end{figure*}
%%%%%%%%%%%%%%%%%%%%%%%%%%%%%%%%%%%%%%%%%%%%%

%%%%%%%%%%%%%%%%%%%%%%%%%%%%%%%%%%%%%%%%%%%%%
\section{Problem Formulation}
\label{sec:problem-formulation}
Let $\mathcal{Q}$ denote the set of admissible natural-language MLOps requirements, and let $q \in \mathcal{Q}$ denote the requirement for a single framework execution. The objective is to transform $q$ into a verified repository and operational cloud deployment while permitting consequential lifecycle transitions only when supported by machine-checkable evidence. Let $\mathcal{S}$ denote the lifecycle-state space and $\mathcal{E}$ the set of controlled transitions in the execution graph, with $\mathcal{E}(s)\subseteq\mathcal{E}$ denoting the transitions that may be considered from lifecycle state $s$. At logical orchestration step $t$, the framework state is

\vspace{-3mm}
\begin{equation}
X_t =
\left(
s_t,
R_t,
D_t,
Z_t,
B_t
\right)
\in
\mathcal{S}\times\mathcal{R}\times\mathcal{D}_{\bot}
\times\mathcal{Z}\times\mathcal{B},
\label{eq:framework-state}
\end{equation}
where $s_t$ is the current lifecycle state, $R_t$ is the repository state, $D_t$ is the observed deployment state, $Z_t$ is the available machine-checkable evidence, and $B_t$ is the retry-budget assignment. Here, $\mathcal{R}$ and $\mathcal{Z}$ denote the repository-state and evidence spaces, respectively, while $\mathcal{D}_{\bot}=\mathcal{D}\cup{\bot}$ denotes the deployment-state space augmented with $\bot$, indicating that no deployment has yet been established. Evidence $Z_t$ may include repository-validation results, test outcomes, security and policy verification results, model-evaluation results, artifact provenance, controlled-execution traces, deployment-verification results, and runtime telemetry. The retry assignment $B_t:\mathcal{E}\rightarrow\mathbb{N}_0$ associates each transition with its remaining retry budget, where $\mathbb{N}_0$ denotes the non-negative integers. Each transition $e\in\mathcal{E}$ is associated with a verification predicate $\phi_e:\mathcal{X}\rightarrow{0,1}$, where $\phi_e(X_t)=1$ indicates that the state and evidence required for that transition satisfy its verification condition. We use $\Phi_{\mathrm{repo}}$, $\Phi_{\mathrm{release}}$, and $\Phi_{\mathrm{runtime}}$ as named predicates for repository and artifact acceptance, controlled cloud release, and runtime acceptability, respectively. These predicates cover the corresponding repository, execution, release, deployment, and runtime evidence required by the lifecycle. Thus, lifecycle progression depends on verified system state rather than agent-reported completion. For an enabled transition $e\in\mathcal{E}(s_t)$, the Graph Orchestrator applies

\vspace{-2mm}
\begin{equation}
T(X_t,e)
=
\begin{cases}
X_{t+1},
&
\phi_e(X_t)=1,
\\[1mm]
X_{t+1}^{\mathrm{corr}},
&
\phi_e(X_t)=0 \land B_t(e)>0,
\\[1mm]
X_{t+1}^{\mathrm{fail}},
&
\phi_e(X_t)=0 \land B_t(e)=0.
\end{cases}
\label{eq:transition}
\end{equation}

where $X_{t+1}$ denotes verified lifecycle progression, $X_{t+1}^{\mathrm{corr}}$ denotes entry into a graph-controlled corrective path, and $X_{t+1}^{\mathrm{fail}}$ denotes a terminal failure state. When correction is entered for transition $e$, its remaining retry budget is decremented as $B_{t+1}(e)=B_t(e)-1$; non-retryable transitions are assigned zero retry budget. For a recoverable verification failure, let $k$ denote the local correction attempt and $\delta_k\in\Delta$ its structured diagnosis, where $\Delta$ is the diagnosis space. The Reflection Agent and Repair Agent operate as $\delta_k = \operatorname{Reflect} \left(R^{(k)},D^{(k)},Z^{(k)}\right)$, $R^{(k+1)} =
\operatorname{Repair} \left(R^{(k)},\delta_k\right), $ where $R^{(k)}$, $D^{(k)}$, and $Z^{(k)}$ denote the repository, deployment, and evidence states available at correction attempt $k$. The corrected repository re-enters the appropriate graph-controlled review, execution, and verification stages, producing fresh evidence $Z^{(k+1)}$. Correction continues only while the relevant verification predicate remains unsatisfied and retry budget remains; otherwise, the Graph Orchestrator follows the configured failure path. The same bounded mechanism applies to runtime failure, drift, degradation, or policy violation and may trigger reflection and repair, adaptation, recovery, or rollback followed by re-verification. Each framework execution therefore terminates in either a verified operational cloud deployment or a terminal failure state with an auditable reason.
%%%%%%%%%%%%%%%%%%%%%%%%%%%%%%%%%%%%%%%%%%%%%

%%%%%%%%%%%%%%%%%%%%%%%%%%%%%%%%%%%%%%%%%%%%%
\section{Overall Framework}
\label{sec:overall-architecture}
\vspace{-2mm}
Figure~\ref{fig:multi-agent-mlops-architecture} presents the agent-level realization of the lifecycle in Figure~\ref{fig:agentic-mlops-workflow}. The platform-independent graph, loop, and agent-harness abstractions are concretely realized on Google Cloud. The framework uses the Google Agent Development Kit (ADK)~\citep{google2026adk}, with the Graph Orchestrator and specialized agents running as containerized workloads on Google Kubernetes Engine (GKE)~\citep{google2026adkgke}. Let $
\mathcal{A}
=
\left\{
A_{\mathrm{gen}},
A_{\mathrm{review}},
A_{\mathrm{exec}},
A_{\mathrm{verify}},
A_{\mathrm{reflect}},
A_{\mathrm{repair}},
A_{\mathrm{release}},
A_{\mathrm{monitor}}
\right\}
$ denote the specialized agents for repository generation, repository review, sandbox execution, verification, reflection, repair, release, and monitoring, respectively. The Graph Orchestrator, denoted by $O_{\mathrm{graph}}$, coordinates these agents according to the execution state and evidence-gated transition rules defined in Section~\ref{sec:problem-formulation}. The primary agent invocation path is 
$
A_{\mathrm{gen}}
\rightarrow
A_{\mathrm{review}}
\rightarrow
A_{\mathrm{exec}}
\rightarrow
A_{\mathrm{verify}}
\rightarrow
A_{\mathrm{release}}
\rightarrow
A_{\mathrm{verify}}
\rightarrow
A_{\mathrm{release}}
\rightarrow
A_{\mathrm{monitor}}.
$. The first invocation of $A_{\mathrm{verify}}$ performs repository artifact verification, whereas the second performs live-cloud verification; similarly, the first invocation of $A_{\mathrm{release}}$ performs build and staging release, whereas the second performs production promotion. Verification failures enter the bounded correction path through $A_{\mathrm{reflect}}$ for diagnosis and $A_{\mathrm{repair}}$ for repository artifact correction. Corrected artifacts re-enter the appropriate graph-controlled review, execution, and verification stages, while failures requiring broader reconstruction may return to $A_{\mathrm{gen}}$. Following repository artifact verification, $A_{\mathrm{release}}$ initiates the controlled release path. Cloud Build performs build and software-supply-chain operations, including source and container scanning, Software Bill of Materials (SBOM) generation, signing, and provenance generation. Artifact Registry stores immutable release artifacts, and Argo CD reconciles the desired deployment state to GKE. OPA/Gatekeeper enforces deployment Policy as Code, while Binary Authorization enforces artifact-attestation requirements before workload admission. Artifacts satisfying these controls are deployed to staging, subjected to live-cloud verification, and promoted to production only when the corresponding verification predicates are satisfied. Figure~\ref{fig:agentic-mlops-deployment-architecture} presents the concrete Google Cloud realization. Agent2Agent (A2A) supports delegation, messages, and artifact references between the Graph Orchestrator and specialized agents, while the Model Context Protocol (MCP) mediates policy-scoped capability access through three paths. \emph{Repository / Browser MCP} provides access to the VS Code and Chrome Sandboxes for repository generation, review, reflection, and repair. \emph{Execution MCP} provides access to the GKE Sandbox with gVisor for controlled artifact execution~\citep{google2026gkeagentsandbox}. \emph{Cloud / Release MCP} provides access to controlled Google Cloud infrastructure for live-cloud verification, release, monitoring, recovery, and rollback. The agent harness applies identity, authorization, network, credential, state, and observability controls across these execution paths. IAM and Workload Identity Federation provide cloud identity and federation; SPIFFE/SPIRE and mutual TLS (mTLS) establish workload identity and authenticated service communication; Role-Based Access Control (RBAC) and Attribute-Based Access Control (ABAC) restrict agent capabilities; Cilium and Kubernetes NetworkPolicy constrain network communication; and Secret Manager controls credential and secret access. AlloyDB for PostgreSQL persists workflow and audit state with row-level security (RLS) for multi-tenant isolation, while Cloud Storage retains execution evidence and provenance. Following production promotion, $A_{\mathrm{monitor}}$ receives runtime evidence from OpenTelemetry, Managed Prometheus, Cloud Logging and Trace, deployment-health telemetry, drift and degradation indicators, and Falco runtime-security alerts. The graph-controlled runtime verification path evaluates this evidence and activates bounded adaptation, recovery, rollback, re-verification, or termination when the corresponding runtime predicates are not satisfied.
%%%%%%%%%%%%%%%%%%%%%%%%%%%%%%%%%%%%%%%%%%%%%

%%%%%%%%%%%%%%%%%%%%%%%%%%%%%%%%%%%%%%%%%%%%%%%%%%%%%%%
\section{Experiments}
\vspace{-2mm}
%%%%%%%%%%%%%%%%%%%%%%%%%%%%%%%%%%%%%%%%%%%%%%%%%%%%%%%
\subsection{Experimental Methodology}
\label{sec:experimentalmethodology}
\vspace{-2mm}
We evaluate the framework through five research questions covering repository completeness and acceptance (RQ1), controlled artifact execution (RQ2), evidence-gated lifecycle progression (RQ3), cloud release and promotion (RQ4), and bounded recovery and termination (RQ5).  \textbf{(a) RQ1: Repository Completeness and Acceptance.} Does the framework generate a complete deployable repository containing the required application code, pipeline, testing, container, CI/CD, infrastructure, Helm, Kubernetes, GitOps, security, provenance, observability, retraining, recovery, and rollback artifacts, and does the resulting repository satisfy the repository-verification predicate? \textbf{(b) RQ2: Controlled Artifact Execution.} Are executable repository artifacts processed through the GKE Sandbox with gVisor, with machine-checkable execution evidence produced and retained for verification? \textbf{(c) RQ3: Evidence-Gated Lifecycle Progression.} Does the Graph Orchestrator permit consequential forward lifecycle transitions only when their required verification predicates are satisfied and block unsupported progression when those predicates are unsatisfied? \textbf{(d) RQ4: Cloud Release and Promotion.} Do repository-verified artifacts progress through the software supply chain, artifact publication, staging deployment, cloud verification, and production promotion, with promotion permitted only after cloud verification succeeds? \textbf{(e) RQ5: Bounded Recovery and Termination.} Do repository, cloud, and runtime verification failures activate the appropriate graph-controlled correction, adaptation, recovery, or rollback path, require re-verification before further progression, and either resume progression within the applicable retry budget or terminate in an auditable failure when that budget is exhausted?
%%%%%%%%%%%%%%%%%%%%%%%%%%%%%%%%%%%%%%%%%%%%%%%%%%%%%%%

%%%%%%%%%%%%%%%%%%%%%%%%%%%%%%%%%%%%%%%%%%%%%%%%%%%%%%%
\subsection{Experimental Scenarios}
\label{sec:experimentalscenarios}
\vspace{-2mm}
The experimental suite uses a single  nominal scenario, three recoverable verification-perturbation scenarios, and one retry-budget-exhaustion scenario. All scenarios use the same execution graph, verification predicates, transition rules, retry semantics, recovery policies, and cloud configuration. Each perturbation introduces a controlled condition only at the targeted verification point required to exercise the corresponding lifecycle path. \textbf{(a) Nominal execution.} The framework executes without an induced verification failure. Successful execution requires a verified operational cloud deployment. Any non-induced recoverable verification failure must be corrected and successfully re-verified within the applicable retry budget. \textbf{(b) Repository-verification perturbation.} A controlled recoverable condition causing repository verification to fail is induced after controlled artifact execution. Release is blocked until the affected artifacts are repaired or regenerated and the required review, execution, and verification stages succeed. \textbf{(c) Cloud-verification perturbation.} A controlled recoverable condition causing cloud verification to fail is induced after staging deployment. Production promotion is blocked until the applicable correction or recovery path completes and cloud re-verification succeeds. \textbf{(d) Runtime-verification perturbation.} A controlled recoverable runtime condition representing service failure, model drift, performance degradation, or policy violation is induced after production promotion. Continued verified operation requires the applicable adaptation, repair, recovery, or rollback path followed by successful runtime re-verification. \textbf{(e) Retry-budget exhaustion.} Each task is assigned a repository, cloud, or runtime verification target using a fixed model-independent assignment. The corresponding failure-inducing condition is maintained across correction or recovery attempts. Recovery continues while the applicable retry budget remains. If the required verification predicate remains unsatisfied when the budget reaches zero, the Graph Orchestrator blocks the corresponding forward progression and terminates the execution in an auditable terminal failure state.
%%%%%%%%%%%%%%%%%%%%%%%%%%%%%%%%%%%%%%%%%%%%%%%%%%%%%%%

%%%%%%%%%%%%%%%%%%%%%%%%%%%%%%%%%%%%%%%%%%%%%%%%%%%%%%%
\subsection{Evaluation Metrics}
\label{sec:evaluationmetrics}
\vspace{-2mm}
We use one overall metric and metrics corresponding to RQ1--RQ5. All metrics lie in $[0,1]$, with higher values indicating better performance, and are reported separately for each language model and experimental scenario where applicable. Table~\ref{tab:evaluation-metrics-summary} summarizes the metrics and their interpretation.

\begin{table}[ht!]
\centering
\vspace{-2mm}
\caption{Summary of evaluation metrics.}
\label{tab:evaluation-metrics-summary}
\small
\renewcommand{\arraystretch}{1.15}
\setlength{\tabcolsep}{4pt}

\begin{tabularx}{\textwidth}{
    >{\raggedright\arraybackslash}p{1.35cm}
    >{\raggedright\arraybackslash}p{4.0cm}
    >{\raggedright\arraybackslash}X
}
\toprule
\textbf{Overall / RQ} & \textbf{Metric} & \textbf{Interpretation} \\
\midrule

\textbf{Overall}
&
\textbf{Verified Operational Deployment Rate (VODR)}
&
Did the entire execution end in a verified operational cloud deployment, with
$\Phi_{\mathrm{repo}}=1$, $\Phi_{\mathrm{release}}=1$, and
$\Phi_{\mathrm{runtime}}=1$?
\\

\textbf{RQ1}
&
\textbf{Repository Completeness Score (RCS)}
&
How complete is the generated repository in terms of the required artifact categories?
\\

\textbf{RQ1}
&
\textbf{Repository Acceptance Rate (RAR)}
&
Did the repository ultimately satisfy $\Phi_{\mathrm{repo}}$ after bounded correction and re-verification?
\\

\textbf{RQ2}
&
\textbf{Controlled Execution Rate (CER)}
&
Were executable repository artifacts processed through the GKE Sandbox with gVisor and accompanied by retained machine-checkable execution evidence?
\\

\textbf{RQ3}
&
\textbf{Verified Progression Rate (VPR)}
&
When a forward transition's verification predicate was satisfied, did the Graph Orchestrator permit progression?
\\

\textbf{RQ3}
&
\textbf{Blocked Forward Progression Rate (BFR)}
&
When a forward transition's verification predicate was unsatisfied, did the Graph Orchestrator block progression?
\\

\textbf{RQ4}
&
\textbf{Release and Promotion Success Rate (RPSR)}
&
After repository verification succeeded, did the execution complete the controlled release path through staging, cloud verification, and production promotion?
\\

\textbf{RQ5}
&
\textbf{Recovery Success Rate (RSR)}
&
After a recoverable verification failure, did the framework recover, re-verify the affected predicate within the retry budget, and resume lifecycle progression?
\\

\textbf{RQ5}
&
\textbf{Exhaustion Termination Rate (ETR)}
&
When the targeted predicate remained unsatisfied through the retry budget, did the framework block progression and terminate in an auditable terminal failure state?
\\

\bottomrule
\end{tabularx}
\vspace{-3mm}
\end{table}
%%%%%%%%%%%%%%%%%%%%%%%%%%%%%%%%%%%%%%%%%%%%%%%%%%%%%%%

%%%%%%%%%%%%%%%%%%%%%%%%%%%%%%%%%%%%%%%%%%%%%%%%%%%%%%%
\subsection{Experimental Setup}
\label{sec:experimentalsetup}
\vspace{-2mm}
We evaluate the framework on a custom benchmark of 100 distinct natural-language cloud MLOps engineering tasks using the Google Cloud Platform (GCP) configuration described in Section~\ref{sec:overall-architecture}.  The representative tasks and benchmark details are provided in the Technical Appendix (Section~\ref{sec:representativetasks}). Existing benchmarks cover ML related software-engineering tasks, but not end-to-end autonomous cloud MLOps. We therefore construct a 100-task benchmark spanning diverse datasets, applications, modalities, and ML workloads.
We compare four language-model backbones: Gemini 2.5 Pro, Gemini 2.5 Flash, Gemini 2.5 Flash-Lite, and GPT-5.6 Sol. Within each execution, the same model is used for all agent functions requiring language-model inference. Each task is evaluated with every model under the five experimental scenarios in Section~\ref{sec:experimentalscenarios}, resulting in $100 \times 4 \times 5 = 2{,}000$ framework executions. Agent prompts, execution graphs, verification predicates, recovery policies, framework implementation, and cloud configuration are held fixed across models. For each experimental scenario, the corresponding perturbation specification is also held fixed across models. No model-specific prompt tuning or verification logic is used. Each retryable transition permits at most 20 correction or recovery attempts. Each execution is additionally limited to 120 minutes of wall-clock time, 10,000,000 aggregate model tokens, and 500 tool calls. We record verification outcomes, lifecycle transitions, correction and recovery attempts, tool calls, token usage, execution time, retained evidence, and terminal state. Google Cloud infrastructure and Vertex AI Gemini inference were covered by eligible multiple Google Cloud free-tier credits. Thus the direct experimental expenditure was therefore limited to GPT-5.6 Sol inference through the OpenAI API.
%%%%%%%%%%%%%%%%%%%%%%%%%%%%%%%%%%%%%%%%%%%%%%%%%%%%%%%

%%%%%%%%%%%%%%%%%%%%%%%%%%%%%%%%%%%%%%%%%%%%%%%%%%%%%%%
\subsection{Results and Analysis}
\label{sec:resultsanalysis}
\vspace{-2mm}
Experimental results are reported for GPT-5.6 Sol, Gemini 2.5 Pro, Gemini 2.5 Flash, and Gemini 2.5 Flash-Lite across the experimental scenarios described in Section~\ref{sec:experimentalscenarios}. Each model--scenario pair comprises 100 executions. VODR, RCS, and RAR are computed over all executions in the scenarios for which each metric is defined; CER is evaluated over executions that reach controlled artifact execution, and RPSR over those that satisfy repository verification. RSR is measured over recoverable verification failures, whereas ETR is measured over retry-budget-exhaustion executions. VPR and BFR are calculated over lifecycle-gate evaluations for which the relevant verification predicate is satisfied and unsatisfied, respectively. The Repository Verification, Cloud Verification, and Runtime Verification columns denote perturbation scenarios with induced failures at the lifecycle gates governed by $\Phi_{\mathrm{repo}}$, $\Phi_{\mathrm{release}}$, and $\Phi_{\mathrm{runtime}}$, respectively. Table~\ref{tab:vodr-results} reports VODR, the fraction of executions whose terminal state satisfies $\Phi_{\mathrm{repo}}$, $\Phi_{\mathrm{release}}$, and $\Phi_{\mathrm{runtime}}$. GPT-5.6 Sol achieves the highest VODR in every reported scenario, followed consistently by Gemini 2.5 Pro, Gemini 2.5 Flash, and Gemini 2.5 Flash-Lite. For all models, VODR decreases under each perturbation relative to nominal execution, consistent with the additional correction or recovery and re-verification required after an induced failure. The framework must produce fresh evidence and re-satisfy the affected verification predicate before lifecycle progression or continued verified operation can resume. RQ5 examines this behavior through RSR, which measures successful recovery within the retry budget, and ETR, which measures correct termination after budget exhaustion. VODR is not reported for retry-budget exhaustion because the failure condition is intentionally maintained until the budget is exhausted, making auditable terminal failure the expected outcome. The remaining results are presented and discussed in the Technical Appendix.

\vspace{-3mm}
\begingroup
\renewcommand{\arraystretch}{1.15}
\setlength{\tabcolsep}{5pt}
\begin{table*}[ht!]
\centering
\caption{Verified Operational Deployment Rate (VODR) across experimental scenarios.}
\label{tab:vodr-results}
\small
\begin{tabular*}{\textwidth}{@{\extracolsep{\fill}}lcccc@{}}
\toprule
\textbf{Model}
& \textbf{Nominal}
& \shortstack{\textbf{Repository}\\\textbf{Verification}}
& \shortstack{\textbf{Cloud}\\\textbf{Verification}}
& \shortstack{\textbf{Runtime}\\\textbf{Verification}} \\
\midrule
GPT-5.6 Sol            & \textbf{0.99} & \textbf{0.95} & \textbf{0.96} & \textbf{0.94} \\
Gemini 2.5 Pro         & 0.78 & 0.62 & 0.65 & 0.59 \\
Gemini 2.5 Flash       & 0.52 & 0.38 & 0.41 & 0.35 \\
Gemini 2.5 Flash-Lite  & 0.38 & 0.24 & 0.27 & 0.21 \\
\bottomrule
\end{tabular*}
\end{table*}
\vspace{-1mm}
\endgroup

\textbf{The key result (see  (able~\ref{tab:rq5-results}) is that the framework enables effective recovery from repository, cloud, and runtime failures while reliably enforcing termination when recovery is exhausted, demonstrating robust bounded autonomy in real-world agentic workflows. These results support our framework’s central novelty by decoupling model-independent verification enforcement from model-dependent recovery capability, thereby enabling bounded, evidence-gated recovery for long-horizon agentic workflows.}

%%%%%%%%%%%%%%%%%%%%%%%%%%%%%%%%%%%%%%%%%%%%%%%%%%%%%%%

\vspace{-2mm}
\section{Conclusion}
\label{sec:conclusion}
\vspace{-3mm}
We presented an evidence-gated multi-agent framework for autonomous cloud MLOps that combines graph-controlled lifecycle orchestration, bounded correction and recovery, and controlled agent execution. The experimental results support the central hypothesis that multi-step, long-horizon MLOps workflows can be governed by machine-checkable evidence: verified transitions proceed, unsupported transitions are blocked, and recoverable or unresolved failures are handled through bounded re-verification or auditable termination. The current evaluation focuses on a representative frontend--backend cloud application rather than a complete production-grade cloud stack. Future work will extend the framework to cover Domain Name System (DNS) configuration and custom domains, external HTTPS load balancing and Transport Layer Security (TLS), Web Application Firewall (WAF) protection, Content Delivery Network (CDN)-backed frontend delivery, distributed rate limiting, and more comprehensive audit analytics. Further evaluation across application architectures, workloads, cloud platforms, and failure modes is needed to assess the generalizability of the framework.
%%%%%%%%%%%%%%%%%%%%%%%%%%%%%%%%%%%%%%%%%%%%%

\clearpage
\newpage

%====================================%
% References 
%====================================% 
\nocite{*}
\bibliographystyle{plainnat}
\bibliography{references}

%====================================%
% Technical Appendix 
%====================================%
\clearpage 

\section{Technical Appendix}
\label{sec:technical_appendix}

%%%%%%%%%%%%%%%%%%%%%%%%%%%%%%%%%%%%%%%%%%%%%%%%%%%%%%%
\subsection{Additional Results}
\label{sec:additionalresults}

\paragraph{(a) RQ1: Repository Completeness and Acceptance.}
Table~\ref{tab:rq1-results} reports the Repository Completeness Score (RCS) and Repository Acceptance Rate (RAR). RCS remains high across models and scenarios, whereas RAR varies substantially more, showing that structural repository completeness does not by itself imply satisfaction of $\Phi_{\mathrm{repo}}$. GPT-5.6 Sol achieves the highest RCS and RAR in all reported scenarios, while the larger RCS--RAR gaps for the Gemini models indicate that artifact-category coverage and repository acceptance remain distinct properties.

\begin{table*}[ht!]
\centering
\caption{Repository completeness and acceptance across experimental scenarios.}
\label{tab:rq1-results}
\small
\renewcommand{\arraystretch}{1.13}
\setlength{\tabcolsep}{4pt}
\begin{tabular*}{\textwidth}{@{\extracolsep{\fill}}llccccc@{}}
\toprule
\textbf{Model}
& \textbf{Metric}
& \textbf{Nominal}
& \shortstack{\textbf{Repository}\\\textbf{Verification}}
& \shortstack{\textbf{Cloud}\\\textbf{Verification}}
& \shortstack{\textbf{Runtime}\\\textbf{Verification}}
& \shortstack{\textbf{Retry-Budget}\\\textbf{Exhaustion}} \\
\midrule
GPT-5.6 Sol
& RCS & \textbf{0.99} & \textbf{0.99} & \textbf{0.99} & \textbf{0.99} & \textbf{0.98} \\
& RAR & \textbf{0.99} & \textbf{0.97} & \textbf{0.99} & \textbf{0.99} & \textbf{0.56} \\
\addlinespace[2pt]
Gemini 2.5 Pro
& RCS & 0.97 & 0.95 & 0.96 & 0.96 & 0.94 \\
& RAR & 0.82 & 0.67 & 0.79 & 0.77 & 0.42 \\
\addlinespace[2pt]
Gemini 2.5 Flash
& RCS & 0.93 & 0.91 & 0.92 & 0.92 & 0.90 \\
& RAR & 0.62 & 0.47 & 0.60 & 0.57 & 0.36 \\
\addlinespace[2pt]
Gemini 2.5 Flash-Lite
& RCS & 0.89 & 0.86 & 0.88 & 0.87 & 0.85 \\
& RAR & 0.46 & 0.30 & 0.43 & 0.39 & 0.28 \\
\bottomrule
\end{tabular*}
\end{table*}

\paragraph{(b) RQ2: Controlled Artifact Execution.}
Table~\ref{tab:rq2-results} reports the Controlled Execution Rate (CER). Among executions that reach controlled artifact execution, CER is 1.00 in all reported cases, indicating that executable repository artifacts are processed through the GKE Sandbox with gVisor and accompanied by retained machine-checkable execution evidence.

\begin{table*}[ht!]
\centering
\caption{Controlled Execution Rate (CER) across experimental scenarios.}
\label{tab:rq2-results}
\small
\renewcommand{\arraystretch}{1.15}
\setlength{\tabcolsep}{4pt}
\begin{tabular*}{\textwidth}{@{\extracolsep{\fill}}lccccc@{}}
\toprule
\textbf{Model}
& \textbf{Nominal}
& \shortstack{\textbf{Repository}\\\textbf{Verification}}
& \shortstack{\textbf{Cloud}\\\textbf{Verification}}
& \shortstack{\textbf{Runtime}\\\textbf{Verification}}
& \shortstack{\textbf{Retry-Budget}\\\textbf{Exhaustion}} \\
\midrule
GPT-5.6 Sol           & 1.00 & 1.00 & 1.00 & 1.00 & 1.00 \\
Gemini 2.5 Pro        & 1.00 & 1.00 & 1.00 & 1.00 & 1.00 \\
Gemini 2.5 Flash      & 1.00 & 1.00 & 1.00 & 1.00 & 1.00 \\
Gemini 2.5 Flash-Lite & 1.00 & 1.00 & 1.00 & 1.00 & 1.00 \\
\bottomrule
\end{tabular*}
\end{table*}

\paragraph{(c) RQ3: Evidence-Gated Lifecycle Progression.}
Table~\ref{tab:rq3-results} reports the Verified Progression Rate (VPR) and Blocked Forward Progression Rate (BFR). Both are 1.00 in all reported cases. The forward progression occurs when the applicable verification predicate is satisfied, while progression is blocked when the predicate is unsatisfied. These results show that the Graph Orchestrator consistently enforces the evidence-gated transition rule defined in Section~\ref{sec:problem-formulation}.

\begin{table*}[ht!]
\centering
\caption{Verified Progression Rate (VPR) and Blocked Forward Progression Rate (BFR) across experimental scenarios.}
\label{tab:rq3-results}
\small
\renewcommand{\arraystretch}{1.13}
\setlength{\tabcolsep}{4pt}
\begin{tabular*}{\textwidth}{@{\extracolsep{\fill}}llccccc@{}}
\toprule
\textbf{Model}
& \textbf{Metric}
& \textbf{Nominal}
& \shortstack{\textbf{Repository}\\\textbf{Verification}}
& \shortstack{\textbf{Cloud}\\\textbf{Verification}}
& \shortstack{\textbf{Runtime}\\\textbf{Verification}}
& \shortstack{\textbf{Retry-Budget}\\\textbf{Exhaustion}} \\
\midrule
GPT-5.6 Sol
& VPR & 1.00 & 1.00 & 1.00 & 1.00 & 1.00 \\
& BFR & 1.00 & 1.00 & 1.00 & 1.00 & 1.00 \\
\addlinespace[2pt]
Gemini 2.5 Pro
& VPR & 1.00 & 1.00 & 1.00 & 1.00 & 1.00 \\
& BFR & 1.00 & 1.00 & 1.00 & 1.00 & 1.00 \\
\addlinespace[2pt]
Gemini 2.5 Flash
& VPR & 1.00 & 1.00 & 1.00 & 1.00 & 1.00 \\
& BFR & 1.00 & 1.00 & 1.00 & 1.00 & 1.00 \\
\addlinespace[2pt]
Gemini 2.5 Flash-Lite
& VPR & 1.00 & 1.00 & 1.00 & 1.00 & 1.00 \\
& BFR & 1.00 & 1.00 & 1.00 & 1.00 & 1.00 \\
\bottomrule
\end{tabular*}
\end{table*}

\paragraph{(d) RQ4: Cloud Release and Promotion.}
Table~\ref{tab:rq4-results} reports the Release and Promotion Success Rate (RPSR), computed over executions that satisfy repository verification. GPT-5.6 Sol achieves the highest RPSR in every reported scenario. The cloud-verification perturbation produces the largest reduction for the Gemini models, consistent with its direct targeting of a verification gate within the release path. The runtime perturbation is introduced only after production promotion; therefore, RPSR in that scenario reflects release outcomes obtained before the induced runtime failure and should not be interpreted as an effect of the runtime perturbation itself.

\begin{table*}[ht!]
\centering
\caption{Release and Promotion Success Rate (RPSR) across experimental scenarios.}
\label{tab:rq4-results}
\small
\renewcommand{\arraystretch}{1.15}
\setlength{\tabcolsep}{5pt}
\begin{tabular*}{\textwidth}{@{\extracolsep{\fill}}lcccc@{}}
\toprule
\textbf{Model}
& \textbf{Nominal}
& \shortstack{\textbf{Repository}\\\textbf{Verification}}
& \shortstack{\textbf{Cloud}\\\textbf{Verification}}
& \shortstack{\textbf{Runtime}\\\textbf{Verification}} \\
\midrule
GPT-5.6 Sol           & \textbf{1.00} & \textbf{0.99} & \textbf{0.98} & \textbf{0.98} \\
Gemini 2.5 Pro        & 0.98 & 0.96 & 0.84 & 0.91 \\
Gemini 2.5 Flash      & 0.90 & 0.87 & 0.70 & 0.79 \\
Gemini 2.5 Flash-Lite & 0.89 & 0.87 & 0.65 & 0.77 \\
\bottomrule
\end{tabular*}
\end{table*}

\paragraph{(e) RQ5: Bounded Recovery and Termination.}
Table~\ref{tab:rq5-results} reports Recovery Success Rate (RSR) for the three recoverable verification targets and Exhaustion Termination Rate (ETR) for retry-budget exhaustion. GPT-5.6 Sol achieves the highest RSR at each recoverable target. For the Gemini models, RSR is lowest for repository-verification failures, while GPT-5.6 Sol remains near ceiling across all three targets. This pattern is consistent with the broader artifact surface covered by $\Phi_{\mathrm{repo}}$, although the results do not isolate the source of the recovery difficulty. For executions that reach the assigned retry-exhaustion target, ETR is 1.00 for every model, showing that the Graph Orchestrator blocks forward progression and terminates when the targeted predicate remains unsatisfied after exhaustion of the applicable retry budget. Together, RSR and ETR show that model capability primarily affects successful recovery within the retry budget, whereas retry-budget exhaustion is handled consistently by the Graph Orchestrator.

\begin{table*}[ht!]
\centering
\caption{Recovery and retry-budget-exhaustion outcomes by verification target.}
\label{tab:rq5-results}
\small
\renewcommand{\arraystretch}{1.13}
\setlength{\tabcolsep}{5pt}
\begin{tabular*}{\textwidth}{@{\extracolsep{\fill}}llcccc@{}}
\toprule
\textbf{Verification Target}
& \textbf{Measure}
& \shortstack{\textbf{GPT-5.6}\\\textbf{Sol}}
& \shortstack{\textbf{Gemini}\\\textbf{2.5 Pro}}
& \shortstack{\textbf{Gemini}\\\textbf{2.5 Flash}}
& \shortstack{\textbf{Gemini}\\\textbf{2.5 Flash-Lite}} \\
\midrule
Repository Verification
& Reached & 99 & 96 & 94 & 88 \\
& RSR     & \textbf{0.98} & 0.70 & 0.50 & 0.34 \\
\addlinespace[2pt]
Cloud Verification
& Reached & 99 & 77 & 57 & 40 \\
& RSR     & \textbf{0.98} & 0.86 & 0.74 & 0.70 \\
\addlinespace[2pt]
Runtime Verification
& Reached & 97 & 70 & 45 & 30 \\
& RSR     & \textbf{0.97} & 0.84 & 0.78 & 0.70 \\
\addlinespace[2pt]
Retry-Budget Exhaustion
& Reached & 94 & 76 & 64 & 55 \\
& ETR     & 1.00 & 1.00 & 1.00 & 1.00 \\
\bottomrule
\end{tabular*}
\end{table*}
%%%%%%%%%%%%%%%%%%%%%%%%%%%%%%%%%%%%%%%%%%%%%%%%%%%%%%%

%%%%%%%%%%%%%%%%%%%%%%%%%%%%%%%%%%%%%%%%%%%%%%%%%%%%%%%
\subsection{Representative Benchmark Tasks}
\label{sec:representativetasks}

The benchmark comprises 100 distinct natural-language cloud MLOps engineering
tasks instantiated from established public datasets and spanning classification,
regression, forecasting, computer vision, natural-language processing, speech,
recommendation, anomaly detection, sensor analytics, and equipment-health
applications. Each task specifies a two-tier application with a frontend for
user interaction and result visualization and a backend providing authenticated
application and inference APIs. Tasks are grouped into \emph{easy},
\emph{medium}, and \emph{hard} tiers according to end-to-end engineering
complexity, including data processing, model workflow, application
functionality, pipeline construction, deployment, monitoring, retraining,
recovery, and rollback. All tasks use the same execution graph, verification
predicates, recovery semantics, and GCP configuration described in
Section~\ref{sec:overall-architecture}. Table~\ref{tab:representative-benchmark-tasks}
shows representative examples. Each task is evaluated unchanged under the five
experimental scenarios in Section~\ref{sec:experimentalscenarios}, with
scenario-specific perturbations introduced only by the experimental procedure.

\begin{table*}[t]
\centering
\caption{Representative tasks from the 100-task cloud MLOps benchmark.}
\label{tab:representative-benchmark-tasks}
\scriptsize
\renewcommand{\arraystretch}{1.06}
\setlength{\tabcolsep}{2.5pt}

\begin{tabularx}{\textwidth}{
    >{\raggedright\arraybackslash}p{0.85cm}
    >{\raggedright\arraybackslash}p{1.85cm}
    >{\raggedright\arraybackslash}p{2.05cm}
    >{\raggedright\arraybackslash}p{6.00cm}
    >{\raggedright\arraybackslash}X
}
\toprule
\textbf{Level}
& \textbf{Dataset}
& \textbf{Application}
& \textbf{Two-Tier Application}
& \textbf{ML Workflow} \\
\midrule

Easy
& MNIST
& Digit classification
& Image-upload frontend with predicted-digit visualization; backend for image
preprocessing and authenticated inference.
& Image classification, reproducible training, evaluation, monitoring, and
retraining. \\

Easy
& California Housing
& House-value regression
& Feature-input frontend with predicted-value visualization; backend for
feature preprocessing and authenticated inference.
& Tabular preprocessing, regression training, evaluation, monitoring, and
retraining. \\

Easy
& IMDB Large Movie Review
& Sentiment classification
& Text-input frontend with sentiment visualization; backend for text
preprocessing and authenticated inference.
& Text preprocessing, sentiment classification, evaluation, monitoring, and
retraining. \\

\midrule

Medium
& CIFAR-10
& Image classification
& Image-upload frontend with class visualization; backend for image
preprocessing and authenticated inference.
& Vision-model training, augmentation, evaluation, serving, monitoring, and
retraining. \\

Medium
& UCI Bike Sharing
& Demand forecasting
& Temporal-input frontend with demand-forecast visualization; backend for
temporal feature processing and authenticated forecasting.
& Temporal feature construction, forecasting, evaluation, monitoring, and
scheduled retraining. \\

Medium
& UCI Human Activity Recognition
& Activity recognition
& Sensor-input frontend with activity visualization; backend for
accelerometer and gyroscope preprocessing and authenticated inference.
& Multivariate sensor preprocessing, activity classification, evaluation,
monitoring, and retraining. \\

Medium
& Speech Commands
& Keyword classification
& Audio-upload frontend with predicted-command visualization; backend for
audio preprocessing, feature extraction, and authenticated inference.
& Audio preprocessing, keyword classification, evaluation, monitoring, and
retraining. \\

\midrule

Hard
& Oxford-IIIT Pet
& Image segmentation
& Image-upload frontend with segmentation-mask visualization; backend for
image preprocessing and authenticated segmentation inference.
& Pixel-level segmentation, training, evaluation, serving, monitoring, and
retraining. \\

Hard
& MovieLens
& Recommendation
& Interactive frontend with ranked recommendations; backend for user--item
processing, ranking, and authenticated recommendation APIs.
& User--item modeling, ranking, offline evaluation, serving, monitoring, and
scheduled retraining. \\

Hard
& NASA C-MAPSS Turbofan Degradation
& Remaining-useful-life estimation
& Equipment-health frontend with remaining-life visualization; backend for
multivariate simulated sensor-sequence processing and authenticated inference.
& Sequence preprocessing, remaining-life estimation, evaluation, degradation
monitoring, and retraining. \\

Hard
& Server Machine Dataset (SMD)
& Anomaly detection
& Time-series frontend with anomaly visualization; backend for multivariate
telemetry processing and authenticated anomaly scoring.
& Multivariate time-series processing, anomaly detection, evaluation, runtime
monitoring, and retraining. \\

\bottomrule
\end{tabularx}
\end{table*}

%%%%%%%%%%%%%%%%%%%%%%%%%%%%%%%%%%%%%%%%%%%%%%%%%%%%%%%

%%%%%%%%%%%%%%%%%%%%%%%%%%%%%%%%%%%%%%%%%%%%%%%%%%%%%%%
\subsection{Ablation Studies}
\label{sec:ablationstudies}
We evaluate GPT-5.6 Sol on the same 100 benchmark tasks to isolate the effects of evidence-gated progression and retry-enabled correction and recovery defined in Section~\ref{sec:problem-formulation}, and to assess sensitivity to the retry budget. The ablations use the repository-, cloud-, and runtime-verification perturbation scenarios in Section~\ref{sec:experimentalscenarios}, which directly exercise the corresponding verification and recovery paths. Unless modified by an ablation, agent prompts, verification predicates, perturbation specifications, recovery policies, cloud configuration, and execution limits remain unchanged from Section~\ref{sec:experimentalsetup}. The Full Framework results in Section~\ref{sec:resultsanalysis} provide the baseline.

\begin{table*}[ht!]
\centering
\caption{Ablation and retry-budget configurations.}
\label{tab:ablation-configurations}
\small
\renewcommand{\arraystretch}{1.13}
\setlength{\tabcolsep}{5pt}

\begin{tabularx}{\textwidth}{
    >{\raggedright\arraybackslash}p{3.2cm}
    >{\raggedright\arraybackslash}X
}
\toprule
\textbf{Variant}
& \textbf{Modification} \\
\midrule

Full Framework
& Evidence-gated transitions, graph-controlled correction and recovery,
re-verification, and retry-budget semantics remain unchanged. \\

Evidence-Gate Bypass
& Verification predicates are evaluated, but their outcomes do not gate
enabled forward transitions. An unsatisfied predicate therefore does not block
progression or enter the associated corrective or terminal-failure path. \\

Zero Retry Budget
& All transitions that are retryable in the Full Framework are assigned zero
retry budget. Evidence gating remains active, but an unsatisfied predicate
causes terminal failure without entering the graph-controlled correction or
recovery path. \\

Retry-Budget Sensitivity
& The retry budget for each retryable transition is varied over
$\{1,3,5,10,20\}$ while evidence gating and all other framework mechanisms
remain unchanged. \\

\bottomrule
\end{tabularx}
\end{table*}

We report VODR, BFR, and RSR using the definitions and evaluation scopes in Section~\ref{sec:evaluationmetrics}. VODR is computed over all executions in each perturbation scenario, BFR over forward-transition evaluations with an unsatisfied verification predicate, and RSR over recoverable verification failures.

\begin{table*}[ht!]
\centering
\caption{Primary ablation results for GPT-5.6 Sol.}
\label{tab:primary-ablation-results}
\scriptsize
\renewcommand{\arraystretch}{1.13}
\setlength{\tabcolsep}{2pt}

\begin{tabular*}{\textwidth}{@{\extracolsep{\fill}}lccccccccc@{}}
\toprule
\textbf{Variant}
& \shortstack{\textbf{VODR}\\\textbf{Repository}}
& \shortstack{\textbf{VODR}\\\textbf{Cloud}}
& \shortstack{\textbf{VODR}\\\textbf{Runtime}}
& \shortstack{\textbf{BFR}\\\textbf{Repository}}
& \shortstack{\textbf{BFR}\\\textbf{Cloud}}
& \shortstack{\textbf{BFR}\\\textbf{Runtime}}
& \shortstack{\textbf{RSR}\\\textbf{Repository}}
& \shortstack{\textbf{RSR}\\\textbf{Cloud}}
& \shortstack{\textbf{RSR}\\\textbf{Runtime}} \\
\midrule

Full Framework
& 0.95
& 0.96
& 0.94
& 1.00
& 1.00
& 1.00
& 0.98
& 0.98
& 0.97 \\

Evidence-Gate Bypass
& 0.00
& 0.00
& 0.00
& 0.00
& 0.00
& 0.00
& 0.00
& 0.00
& 0.00 \\

Zero Retry Budget
& 0.00
& 0.00
& 0.00
& 1.00
& 1.00
& 1.00
& 0.00
& 0.00
& 0.00 \\

\bottomrule
\end{tabular*}
\end{table*}

Evidence-Gate Bypass reduces BFR to $0.00$ because an unsatisfied predicate no longer blocks the corresponding forward transition. Since the associated corrective path is not entered, the induced failure is not corrected and re-verified, yielding RSR of $0.00$. The affected predicate therefore remains unsatisfied at termination, and VODR is $0.00$ in all three perturbation scenarios. Under Zero Retry Budget, evidence gating remains active and BFR remains $1.00$, but an induced verification failure cannot enter a correction or recovery attempt. The execution therefore terminates at the failed gate, yielding RSR and VODR of $0.00$. We next vary the retry budget under the same three recoverable perturbation scenarios. A budget of 20 corresponds to the Full Framework configuration in Section~\ref{sec:experimentalsetup}.

\begin{table*}[ht!]
\centering
\caption{Retry-budget sensitivity for GPT-5.6 Sol.}
\label{tab:retry-budget-sensitivity}
\small
\renewcommand{\arraystretch}{1.13}
\setlength{\tabcolsep}{4pt}

\begin{tabular*}{\textwidth}{@{\extracolsep{\fill}}ccccccc@{}}
\toprule
\shortstack{\textbf{Retry}\\\textbf{Budget}}
& \shortstack{\textbf{VODR}\\\textbf{Repository}}
& \shortstack{\textbf{VODR}\\\textbf{Cloud}}
& \shortstack{\textbf{VODR}\\\textbf{Runtime}}
& \shortstack{\textbf{RSR}\\\textbf{Repository}}
& \shortstack{\textbf{RSR}\\\textbf{Cloud}}
& \shortstack{\textbf{RSR}\\\textbf{Runtime}} \\
\midrule

1
& 0.74
& 0.77
& 0.71
& 0.75
& 0.78
& 0.74 \\

3
& 0.87
& 0.89
& 0.83
& 0.88
& 0.90
& 0.86 \\

5
& 0.93
& 0.94
& 0.89
& 0.94
& 0.95
& 0.92 \\

10
& 0.95
& 0.96
& 0.93
& 0.96
& 0.97
& 0.96 \\

20
& 0.95
& 0.96
& 0.94
& 0.98
& 0.98
& 0.97 \\

\bottomrule
\end{tabular*}
\end{table*}

VODR and RSR are non-decreasing with the retry budget, with the largest gains at smaller budgets. Runtime RSR increases from $0.74$ with one permitted attempt to $0.92$ with five and $0.97$ with twenty, with similar behavior for repository and cloud verification. The retry budget defines the maximum number of correction or recovery attempts; successful re-verification terminates the
corresponding recovery path before the budget is exhausted. Together, the ablations show that evidence gating prevents unsupported forward progression, while a positive retry budget enables bounded correction or recovery before terminal failure. Increasing the retry budget improves recovery and verified operational deployment, with limited additional gains beyond ten attempts.
%%%%%%%%%%%%%%%%%%%%%%%%%%%%%%%%%%%%%%%%%%%%%%%%%%%%%%%

%%%%%%%%%%%%%%%%%%%%%%%%%%%%%%%%%%%%%%%%%%%%%%%%%%%%%%%
\subsection{Lifecycle and State Management}
\label{sec:lifecycle-state-management}
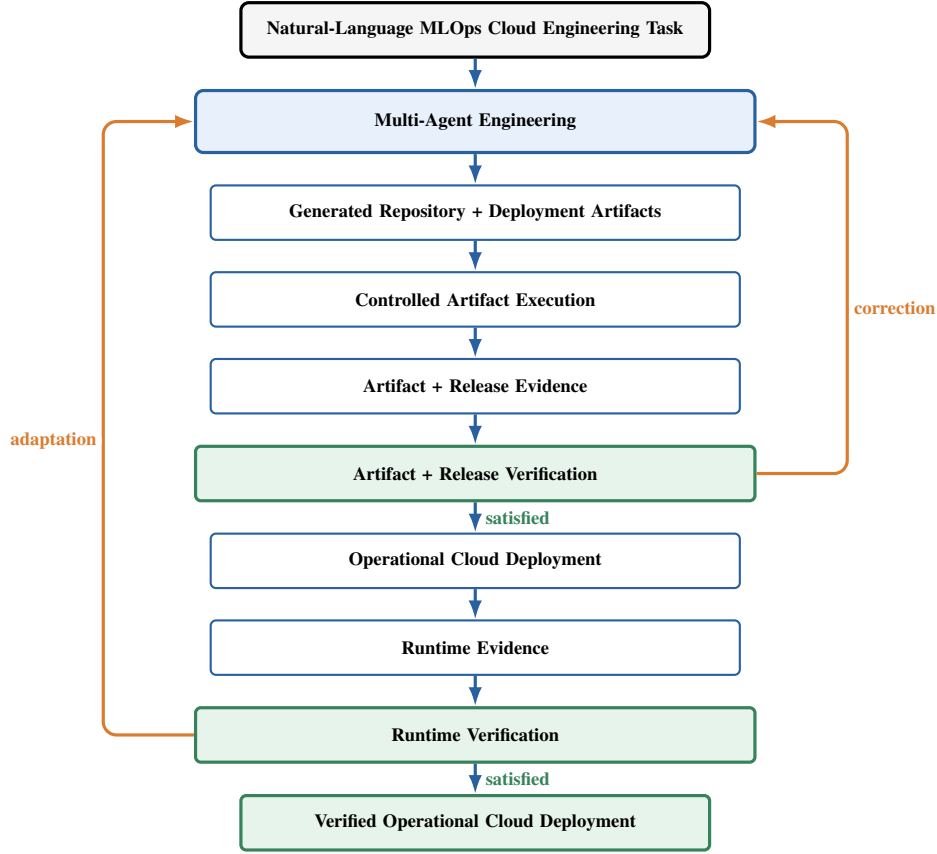
\begin{figure*}[ht!]
\centering

\begin{tikzpicture}[
    node distance=4mm
]

% ==============================================================
% Natural-language task
% ==============================================================
\node[
    request,
    minimum width=6.2cm,
    text width=5.8cm,
    minimum height=0.72cm,
    font=\scriptsize\bfseries
] (request) {
    Natural-Language MLOps Cloud Engineering Task
};

% ==============================================================
% Multi-agent engineering
% ==============================================================
\node[
    graphcontrol,
    below=of request,
    minimum width=7.4cm,
    text width=7.0cm,
    minimum height=0.82cm,
    font=\scriptsize\bfseries
] (engineering) {
    Multi-Agent Engineering
};

% ==============================================================
% Generated repository and deployment artifacts
% ==============================================================
\node[
    process,
    below=of engineering,
    minimum width=7.0cm,
    text width=6.6cm,
    minimum height=0.72cm,
    font=\scriptsize
] (artifacts) {
    \textbf{Generated Repository + Deployment Artifacts}
};

% ==============================================================
% Controlled artifact execution
% ==============================================================
\node[
    process,
    below=of artifacts,
    minimum width=7.0cm,
    text width=6.6cm,
    minimum height=0.72cm,
    font=\scriptsize
] (execution) {
    \textbf{Controlled Artifact Execution}
};

% ==============================================================
% Artifact + release evidence
% ==============================================================
\node[
    process,
    below=of execution,
    minimum width=7.0cm,
    text width=6.6cm,
    minimum height=0.72cm,
    font=\scriptsize
] (releaseevidence) {
    \textbf{Artifact + Release Evidence}
};

% ==============================================================
% Artifact + release verification
% ==============================================================
\node[
    verification,
    below=of releaseevidence,
    minimum width=7.4cm,
    text width=7.0cm,
    minimum height=0.72cm,
    font=\scriptsize\bfseries
] (releaseverification) {
    Artifact + Release Verification
};

% ==============================================================
% Operational cloud deployment
% ==============================================================
\node[
    process,
    below=of releaseverification,
    minimum width=7.0cm,
    text width=6.6cm,
    minimum height=0.72cm,
    font=\scriptsize
] (deployment) {
    \textbf{Operational Cloud Deployment}
};

% ==============================================================
% Runtime evidence
% ==============================================================
\node[
    process,
    below=of deployment,
    minimum width=7.0cm,
    text width=6.6cm,
    minimum height=0.72cm,
    font=\scriptsize
] (runtimeevidence) {
    \textbf{Runtime Evidence}
};

% ==============================================================
% Runtime verification
% ==============================================================
\node[
    verification,
    below=of runtimeevidence,
    minimum width=7.4cm,
    text width=7.0cm,
    minimum height=0.72cm,
    font=\scriptsize\bfseries
] (runtimeverification) {
    Runtime Verification
};

% ==============================================================
% Verified operational cloud deployment
% ==============================================================
\node[
    verification,
    below=of runtimeverification,
    minimum width=6.2cm,
    text width=5.8cm,
    minimum height=0.72cm,
    font=\scriptsize\bfseries
] (completion) {
    Verified Operational Cloud Deployment
};

% ==============================================================
% Forward lifecycle
% ==============================================================

\draw[mainarrow]
    (request) -- (engineering);

\draw[mainarrow]
    (engineering) -- (artifacts);

\draw[mainarrow]
    (artifacts) -- (execution);

\draw[mainarrow]
    (execution) -- (releaseevidence);

\draw[mainarrow]
    (releaseevidence) -- (releaseverification);

\draw[mainarrow]
    (releaseverification)
    -- node[
        right,
        font=\scriptsize\bfseries,
        text=verifygreen
    ] {satisfied}
    (deployment);

\draw[mainarrow]
    (deployment) -- (runtimeevidence);

\draw[mainarrow]
    (runtimeevidence) -- (runtimeverification);

\draw[mainarrow]
    (runtimeverification)
    -- node[
        right,
        font=\scriptsize\bfseries,
        text=verifygreen
    ] {satisfied}
    (completion);

% ==============================================================
% Correction path:
% Artifact + Release Verification -> Multi-Agent Engineering
% ==============================================================

\coordinate (correctionlane)
    at ($(releaseverification.east)+(12mm,0)$);

\coordinate (correctiontop)
    at (correctionlane |- engineering.east);

\draw[looparrow]
    (releaseverification.east)
    -- (correctionlane)
    -- node[
        looplabel,
        right,
        pos=0.47
    ] {correction}
    (correctiontop)
    -- (engineering.east);

% ==============================================================
% Adaptation path:
% Runtime Verification -> Multi-Agent Engineering
% ==============================================================

\coordinate (adaptationlane)
    at ($(runtimeverification.west)+(-12mm,0)$);

\coordinate (adaptationtop)
    at (adaptationlane |- engineering.west);

\draw[looparrow]
    (runtimeverification.west)
    -- (adaptationlane)
    -- node[
        looplabel,
        left,
        pos=0.48
    ] {adaptation}
    (adaptationtop)
    -- (engineering.west);

\end{tikzpicture}

\caption{
High-level lifecycle of the evidence-gated autonomous cloud MLOps framework.
A natural-language MLOps cloud engineering task progresses through multi-agent
engineering, artifact generation and controlled execution, artifact and release
verification, cloud deployment, and runtime verification. Failed artifact or
release verification returns to multi-agent engineering through the correction
path, while runtime verification can return through the adaptation path.
The lifecycle is therefore cyclic rather than a one-shot CI/CD pipeline.
}
\label{fig:agentic-mlops-lifecycle}
\vspace{-3mm}
\end{figure*}

\begin{figure}[ht!]
\centering

\begin{tikzpicture}[
    node distance=4mm
]

\node[
    graphcontrol,
    minimum width=4.8cm,
    text width=4.3cm,
    minimum height=0.8cm,
    font=\scriptsize
] (orchestrator) {
    \textbf{Graph Orchestrator}\\
    workflow and execution state
};

\node[
    process,
    below=of orchestrator,
    minimum width=4.8cm,
    text width=4.3cm,
    minimum height=0.75cm,
    font=\scriptsize
] (alloy) {
    \textbf{AlloyDB for PostgreSQL}\\
    persistent workflow and audit state
};

\node[
    verification,
    below=of alloy,
    minimum width=4.8cm,
    text width=4.3cm,
    minimum height=0.75cm,
    font=\scriptsize
] (rls) {
    \textbf{PostgreSQL RLS}\\
    tenant isolation
};

\draw[mainarrow] (orchestrator) -- (alloy);
\draw[mainarrow] (alloy) -- (rls);

\end{tikzpicture}

\caption{
Persistent workflow and audit state using AlloyDB for PostgreSQL with
row-level security for tenant isolation.
}
\label{fig:persistent-state}
\end{figure}
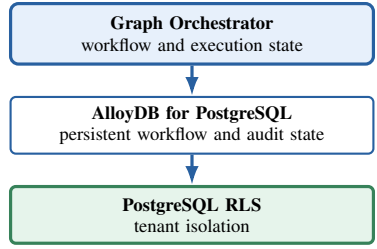

Figure~\ref{fig:agentic-mlops-lifecycle} summarizes the evidence-gated lifecycle from a natural-language MLOps cloud engineering task to a verified operational cloud deployment. Generated repository and deployment artifacts undergo controlled execution, artifact and release verification, deployment, and runtime verification. Predicate satisfaction permits forward progression;
failed artifact or release verification enters the correction path, whereas failed runtime verification enters the adaptation path, with both returning to multi-agent engineering for subsequent re-verification. Persistent workflow and audit state is maintained in AlloyDB for PostgreSQL, with PostgreSQL row-level security providing tenant isolation (Figure~\ref{fig:persistent-state}). The corresponding execution, release,
security, and runtime mechanisms are described in Section~\ref{sec:operationalrealization}.
%%%%%%%%%%%%%%%%%%%%%%%%%%%%%%%%%%%%%%%%%%%%%%%%%%%%%%%

%%%%%%%%%%%%%%%%%%%%%%%%%%%%%%%%%%%%%%%%%%%%%%%%%%%%%%%
%%%%%%%%%%%%%%%%%%%%%%%%%%%%%%%%%%%%%%%%%%%%%%%%%%%%%%%

%%%%%%%%%%%%%%%%%%%%%%%%%%%%%%%%%%%%%%%%%%%%%
%%%%%%%%%%%%%%%%%%%%%%%%%%%%%%%%%%%%%%%%%%%%%

%%%%%%%%%%%%%%%%%%%%%%%%%%%%%%%%%%%%%%%%%%%%%
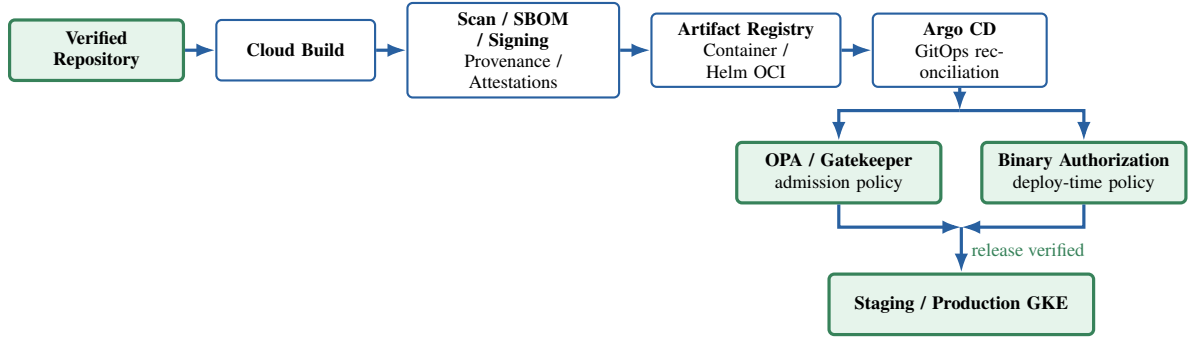
\begin{figure*}[ht!]
\centering

\begin{tikzpicture}[
    node distance=4mm
]

\node[
    verification,
    minimum width=2.3cm,
    text width=1.9cm,
    minimum height=0.8cm,
    font=\scriptsize
] (repo) {
    \textbf{Verified\\Repository}
};

\node[
    process,
    right=of repo,
    minimum width=2.1cm,
    text width=1.7cm,
    minimum height=0.8cm,
    font=\scriptsize
] (build) {
    \textbf{Cloud Build}
};

\node[
    process,
    right=of build,
    minimum width=2.8cm,
    text width=2.4cm,
    minimum height=0.8cm,
    font=\scriptsize
] (supply) {
    \textbf{Scan / SBOM / Signing}\\
    Provenance / Attestations
};

\node[
    process,
    right=of supply,
    minimum width=2.5cm,
    text width=2.1cm,
    minimum height=0.8cm,
    font=\scriptsize
] (registry) {
    \textbf{Artifact Registry}\\
    Container / Helm OCI
};

\node[
    process,
    right=of registry,
    minimum width=2.3cm,
    text width=1.9cm,
    minimum height=0.8cm,
    font=\scriptsize
] (argocd) {
    \textbf{Argo CD}\\
    GitOps reconciliation
};

\draw[mainarrow] (repo) -- (build);
\draw[mainarrow] (build) -- (supply);
\draw[mainarrow] (supply) -- (registry);
\draw[mainarrow] (registry) -- (argocd);

% ------------------------------------------------------------------
% Deployment policy gates
% ------------------------------------------------------------------
\node[
    verification,
    below=7mm of argocd,
    xshift=-1.6cm,
    minimum width=2.7cm,
    text width=2.3cm,
    minimum height=0.8cm,
    font=\scriptsize
] (gatekeeper) {
    \textbf{OPA / Gatekeeper}\\
    admission policy
};

\node[
    verification,
    right=5mm of gatekeeper,
    minimum width=2.7cm,
    text width=2.3cm,
    minimum height=0.8cm,
    font=\scriptsize
] (binauth) {
    \textbf{Binary Authorization}\\
    deploy-time policy
};

\coordinate (policybus)
    at ($(argocd.south)+(0,-3mm)$);

\draw[mainarrow] (argocd.south) -- (policybus);
\draw[mainarrow] (policybus) -| (gatekeeper.north);
\draw[mainarrow] (policybus) -| (binauth.north);

% ------------------------------------------------------------------
% Policy paths rejoin before deployment
% ------------------------------------------------------------------
\coordinate (policyjoin)
    at ($(gatekeeper.south)!0.5!(binauth.south)+(0,-3mm)$);

\draw[mainarrow] (gatekeeper.south) |- (policyjoin);
\draw[mainarrow] (binauth.south) |- (policyjoin);

\node[
    verification,
    below=6mm of policyjoin,
    minimum width=3.5cm,
    text width=3.0cm,
    minimum height=0.8cm,
    font=\scriptsize
] (deploy) {
    \textbf{Staging / Production GKE}
};

\draw[mainarrow]
(policyjoin)
-- node[
    right,
    font=\scriptsize,
    text=verifygreen
] {release verified}
(deploy.north);

\end{tikzpicture}

\caption{
Controlled software-supply-chain and release path through build,
artifact publication, GitOps reconciliation, policy enforcement, and
deployment to GKE.
}
\label{fig:supply-chain-release}
\end{figure*}
%%%%%%%%%%%%%%%%%%%%%%%%%%%%%%%%%%%%%%%%%%%%%

%%%%%%%%%%%%%%%%%%%%%%%%%%%%%%%%%%%%%%%%%%%%%
\begin{figure}[ht!]
\centering

\begin{tikzpicture}[
    node distance=4mm and 4mm
]

\node[
    graphcontrol,
    minimum width=4.8cm,
    text width=4.3cm,
    minimum height=0.8cm,
    font=\scriptsize
] (workloads) {
    \textbf{Agent and Cloud Workloads}
};

\node[
    process,
    below=6mm of workloads,
    xshift=-1.9cm,
    minimum width=3.5cm,
    text width=3.1cm,
    minimum height=0.75cm,
    font=\scriptsize
] (spiffe) {
    \textbf{SPIFFE / SPIRE}\\
    workload identity
};

\node[
    process,
    right=4mm of spiffe,
    minimum width=3.5cm,
    text width=3.1cm,
    minimum height=0.75cm,
    font=\scriptsize
] (mtls) {
    \textbf{mTLS / AuthN / AuthZ}\\
    authenticated access
};

\node[
    process,
    below=4mm of spiffe,
    minimum width=3.5cm,
    text width=3.1cm,
    minimum height=0.75cm,
    font=\scriptsize
] (cilium) {
    \textbf{Cilium / NetworkPolicy}\\
    network isolation
};

\node[
    process,
    right=4mm of cilium,
    minimum width=3.5cm,
    text width=3.1cm,
    minimum height=0.75cm,
    font=\scriptsize
] (secrets) {
    \textbf{Secret Manager}\\
    managed secret access
};

\coordinate (securitybus)
    at ($(workloads.south)+(0,-3mm)$);

\draw[mainarrow] (workloads.south) -- (securitybus);
\draw[mainarrow] (securitybus) -| (spiffe.north);
\draw[mainarrow] (securitybus) -| (mtls.north);

\draw[mainarrow] (spiffe.south) -- (cilium.north);
\draw[mainarrow] (mtls.south) -- (secrets.north);

\end{tikzpicture}

\caption{
Complementary security controls for workload identity, authenticated
communication, network isolation, and managed secret access.
}
\label{fig:security-controls}
\end{figure}
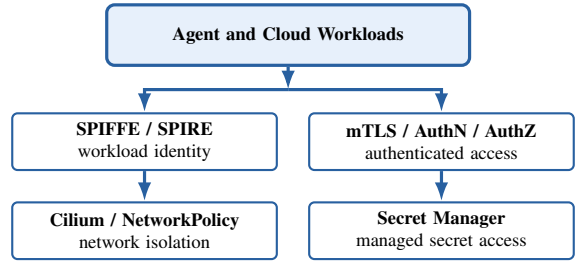
%%%%%%%%%%%%%%%%%%%%%%%%%%%%%%%%%%%%%%%%%%%%%

%%%%%%%%%%%%%%%%%%%%%%%%%%%%%%%%%%%%%%%%%%%%%
\begin{figure*}[ht!]
\centering

\begin{tikzpicture}[
    node distance=4mm
]

% ------------------------------------------------------------------
% Runtime evidence
% ------------------------------------------------------------------
\node[
    process,
    minimum width=1.9cm,
    minimum height=0.65cm,
    font=\scriptsize
] (metrics) {
    \textbf{Metrics}
};

\node[
    process,
    right=of metrics,
    minimum width=1.9cm,
    minimum height=0.65cm,
    font=\scriptsize
] (logs) {
    \textbf{Logs}
};

\node[
    process,
    right=of logs,
    minimum width=1.9cm,
    minimum height=0.65cm,
    font=\scriptsize
] (traces) {
    \textbf{Traces}
};

\node[
    process,
    right=of traces,
    minimum width=2.1cm,
    minimum height=0.65cm,
    font=\scriptsize
] (falco) {
    \textbf{Falco Alerts}
};

% ------------------------------------------------------------------
% Runtime verification
% ------------------------------------------------------------------
\coordinate (evidencebus)
    at ($(logs.south)!0.5!(traces.south)+(0,-4mm)$);

\node[
    verification,
    below=9mm of evidencebus,
    minimum width=3.7cm,
    text width=3.2cm,
    minimum height=0.8cm,
    font=\scriptsize
] (runtime) {
    \textbf{Runtime Verification}
};

\draw[mainarrow] (metrics.south) |- (evidencebus);
\draw[mainarrow] (logs.south) |- (evidencebus);
\draw[mainarrow] (traces.south) |- (evidencebus);
\draw[mainarrow] (falco.south) |- (evidencebus);
\draw[mainarrow] (evidencebus) -- (runtime.north);

% ------------------------------------------------------------------
% Monitoring
% ------------------------------------------------------------------
\node[
    process,
    right=9mm of runtime,
    minimum width=2.8cm,
    text width=2.4cm,
    minimum height=0.8cm,
    font=\scriptsize
] (monitor) {
    \textbf{Monitoring Agent}
};

\draw[mainarrow]
(runtime)
-- node[
    above,
    font=\scriptsize,
    text=verifygreen
] {acceptable}
(monitor);

% ------------------------------------------------------------------
% Bounded adaptation
% ------------------------------------------------------------------
\node[
    process,
    below=8mm of runtime,
    xshift=-1.7cm,
    minimum width=2.8cm,
    text width=2.4cm,
    minimum height=0.8cm,
    font=\scriptsize
] (reflect) {
    \textbf{Reflection Agent}\\
    diagnose
};

\node[
    process,
    right=5mm of reflect,
    minimum width=2.8cm,
    text width=2.4cm,
    minimum height=0.8cm,
    font=\scriptsize
] (repair) {
    \textbf{Repair Agent}\\
    correct artifacts
};

\node[
    graphcontrol,
    right=5mm of repair,
    minimum width=3.2cm,
    text width=2.8cm,
    minimum height=0.8cm,
    font=\scriptsize
] (reentry) {
    \textbf{Lifecycle Re-entry}\\
    verification and release
};

\draw[looparrow]
(runtime.south)
|- node[
    looplabel,
    left
] {adverse evidence}
(reflect.north);

\draw[looparrow]
(reflect.east)
-- node[
    looplabel,
    above
] {diagnosis}
(repair.west);

\draw[looparrow]
(repair.east)
-- node[
    looplabel,
    above
] {correct and retry}
(reentry.west);

\end{tikzpicture}

\caption{
Runtime evidence and bounded adaptation. Metrics, logs, traces, and Falco
alerts support runtime verification, while adverse evidence invokes
reflection and repair before controlled lifecycle re-entry.
}
\label{fig:runtime-evidence}
\end{figure*}
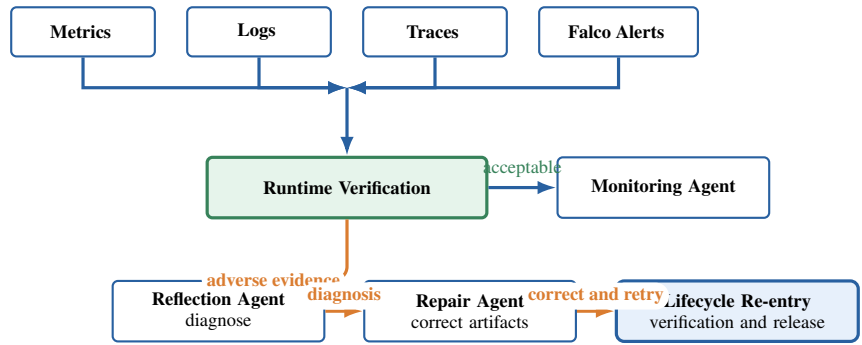
%%%%%%%%%%%%%%%%%%%%%%%%%%%%%%%%%%%%%%%%%%%%%

%%%%%%%%%%%%%%%%%%%%%%%%%%%%%%%%%%%%%%%%%%%%%%%%%%%%%%%
\subsection{Operational Realization}
\label{sec:operationalrealization}
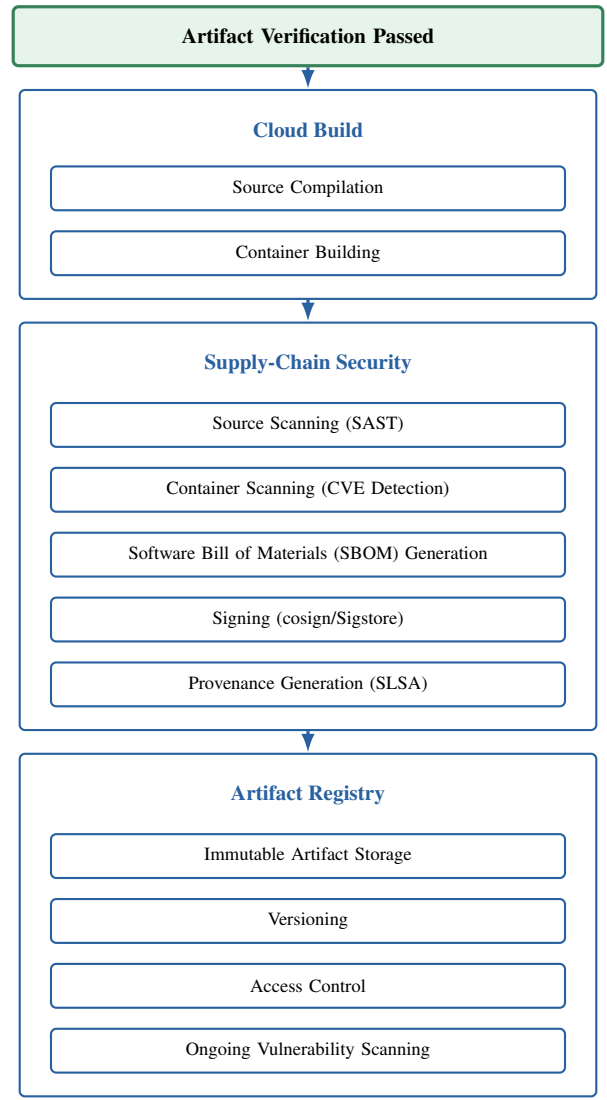
\begin{figure}[t]
\centering

\begin{tikzpicture}[
    node distance=4.5mm,
    substep/.style={
        process,
        minimum width=6.8cm,
        text width=6.35cm,
        minimum height=0.58cm,
        inner ysep=2pt,
        font=\scriptsize,
        align=center
    }
]

% ==================================================================
% Artifact verification
% ==================================================================
\node[
    verification,
    minimum width=7.8cm,
    text width=7.35cm,
    minimum height=0.78cm,
    font=\footnotesize\bfseries,
    align=center
] (artifactverification) {
    Artifact Verification Passed
};

% ==================================================================
% Cloud Build
% ==================================================================

% Title
\node[
    below=6mm of artifactverification,
    font=\footnotesize\bfseries,
    text=graphblue
] (buildtitle) {
    Cloud Build
};

% Sub-boxes
\node[
    substep,
    below=2.5mm of buildtitle
] (compile) {
    Source Compilation
};

\node[
    substep,
    below=2.5mm of compile
] (containerbuild) {
    Container Building
};

% Outer Cloud Build box
\begin{scope}[on background layer]
\node[
    process,
    fit=(buildtitle)(compile)(containerbuild),
    inner xsep=4mm,
    inner ysep=3mm
] (cloudbuild) {};
\end{scope}

% ==================================================================
% Supply-Chain Security
% ==================================================================

% Title
\node[
    below=6mm of cloudbuild,
    font=\footnotesize\bfseries,
    text=graphblue
] (securitytitle) {
    Supply-Chain Security
};

% Sub-boxes
\node[
    substep,
    below=2.5mm of securitytitle
] (sast) {
    Source Scanning (SAST)
};

\node[
    substep,
    below=2.5mm of sast
] (cvescan) {
    Container Scanning (CVE Detection)
};

\node[
    substep,
    below=2.5mm of cvescan
] (sbom) {
    Software Bill of Materials (SBOM) Generation
};

\node[
    substep,
    below=2.5mm of sbom
] (signing) {
    Signing (cosign/Sigstore)
};

\node[
    substep,
    below=2.5mm of signing
] (provenance) {
    Provenance Generation (SLSA)
};

% Outer Supply-Chain Security box
\begin{scope}[on background layer]
\node[
    process,
    fit={
        (securitytitle)
        (sast)
        (cvescan)
        (sbom)
        (signing)
        (provenance)
    },
    inner xsep=4mm,
    inner ysep=3mm
] (supplychain) {};
\end{scope}

% ==================================================================
% Artifact Registry
% ==================================================================

% Title
\node[
    below=6mm of supplychain,
    font=\footnotesize\bfseries,
    text=graphblue
] (registrytitle) {
    Artifact Registry
};

% Sub-boxes
\node[
    substep,
    below=2.5mm of registrytitle
] (immutable) {
    Immutable Artifact Storage
};

\node[
    substep,
    below=2.5mm of immutable
] (versioning) {
    Versioning
};

\node[
    substep,
    below=2.5mm of versioning
] (accesscontrol) {
    Access Control
};

\node[
    substep,
    below=2.5mm of accesscontrol
] (vulnerability) {
    Ongoing Vulnerability Scanning
};

% Outer Artifact Registry box
\begin{scope}[on background layer]
\node[
    process,
    fit={
        (registrytitle)
        (immutable)
        (versioning)
        (accesscontrol)
        (vulnerability)
    },
    inner xsep=4mm,
    inner ysep=3mm
] (artifactregistry) {};
\end{scope}

% ==================================================================
% Forward flow
% ==================================================================
\draw[mainarrow]
    (artifactverification.south) -- (cloudbuild.north);

\draw[mainarrow]
    (cloudbuild.south) -- (supplychain.north);

\draw[mainarrow]
    (supplychain.south) -- (artifactregistry.north);

\end{tikzpicture}

\caption{The figure depicts build and Artifact Pipeline.}
\label{fig:artifact-supply-chain-flow}

\end{figure}

\begin{figure*}[t]
\centering

\begin{tikzpicture}[
    scale=0.94,
    transform shape,
    securitysub/.style={
        process,
        minimum width=7.0cm,
        text width=6.55cm,
        minimum height=0.48cm,
        inner xsep=3pt,
        inner ysep=1pt,
        font=\scriptsize,
        align=center
    }
]

% ==================================================================
% Layer 1: Cloud Identity & Federation
% ==================================================================
\node[
    font=\footnotesize\bfseries,
    text=verifygreen,
    align=center
] (layeronetitle) {
    Layer 1: Cloud Identity \& Federation
};

\node[
    securitysub,
    below=1.0mm of layeronetitle,
    font=\scriptsize\bfseries
] (iamwif) {
    IAM + Workload Identity Federation
};

\node[
    securitysub,
    below=1.0mm of iamwif
] (cloudaccess) {
    Kubernetes Workload Access to Google Cloud Resources
};

\node[
    securitysub,
    below=1.0mm of cloudaccess
] (federatedidentity) {
    Federated Workload Identities
};

\node[
    securitysub,
    below=1.0mm of federatedidentity
] (serviceaccountimpersonation) {
    Optional Service-Account Impersonation
};

\begin{scope}[on background layer]
\node[
    verification,
    fit={
        (layeronetitle)
        (iamwif)
        (cloudaccess)
        (federatedidentity)
        (serviceaccountimpersonation)
    },
    inner xsep=3mm,
    inner ysep=2mm
] (layerone) {};
\end{scope}

% ==================================================================
% Layer 2: Workload Identity & mTLS
% ==================================================================
\node[
    font=\footnotesize\bfseries,
    text=verifygreen,
    below=3.2mm of layerone,
    align=center
] (layertwotitle) {
    Layer 2: Workload Identity \& mTLS
};

\node[
    securitysub,
    below=1.0mm of layertwotitle,
    font=\scriptsize\bfseries
] (spiffespire) {
    SPIFFE / SPIRE + mTLS
};

\node[
    securitysub,
    below=1.0mm of spiffespire
] (spiffeidentity) {
    SPIFFE Workload Identity
};

\node[
    securitysub,
    below=1.0mm of spiffeidentity
] (svid) {
    X.509-SVID / JWT-SVID
};

\node[
    securitysub,
    below=1.0mm of svid
] (serviceauthentication) {
    Service-to-Service Authentication
};

\node[
    securitysub,
    below=1.0mm of serviceauthentication
] (encryptedcommunication) {
    Encrypted Authenticated Communication
};

\begin{scope}[on background layer]
\node[
    verification,
    fit={
        (layertwotitle)
        (spiffespire)
        (spiffeidentity)
        (svid)
        (serviceauthentication)
        (encryptedcommunication)
    },
    inner xsep=3mm,
    inner ysep=2mm
] (layertwo) {};
\end{scope}

% ==================================================================
% Layer 3: Authorization
% ==================================================================
\node[
    font=\footnotesize\bfseries,
    text=verifygreen,
    below=3.2mm of layertwo,
    align=center
] (layerthreetitle) {
    Layer 3: Authorization
};

\node[
    securitysub,
    below=1.0mm of layerthreetitle,
    font=\scriptsize\bfseries
] (rbacabac) {
    RBAC + ABAC
};

\node[
    securitysub,
    below=1.0mm of rbacabac
] (leastprivilege) {
    Least-Privilege Access
};

\node[
    securitysub,
    below=1.0mm of leastprivilege
] (roleauthorization) {
    Role-Based Authorization
};

\node[
    securitysub,
    below=1.0mm of roleauthorization
] (attributepolicies) {
    Attribute-Based Policies
};

\node[
    securitysub,
    below=1.0mm of attributepolicies
] (granularpermissions) {
    Granular Permissions
};

\begin{scope}[on background layer]
\node[
    verification,
    fit={
        (layerthreetitle)
        (rbacabac)
        (leastprivilege)
        (roleauthorization)
        (attributepolicies)
        (granularpermissions)
    },
    inner xsep=3mm,
    inner ysep=2mm
] (layerthree) {};
\end{scope}

% ==================================================================
% Layer 4: Network Isolation
% ==================================================================
\node[
    font=\footnotesize\bfseries,
    text=verifygreen,
    below=3.2mm of layerthree,
    align=center
] (layerfourtitle) {
    Layer 4: Network Isolation
};

\node[
    securitysub,
    below=1.0mm of layerfourtitle,
    font=\scriptsize\bfseries
] (ciliumpolicy) {
    Cilium + Kubernetes NetworkPolicy
};

\node[
    securitysub,
    below=1.0mm of ciliumpolicy
] (podisolation) {
    Pod-to-Pod Isolation
};

\node[
    securitysub,
    below=1.0mm of podisolation
] (networksegmentation) {
    Network Segmentation
};

\node[
    securitysub,
    below=1.0mm of networksegmentation
] (ingressegress) {
    Ingress / Egress Policy
};

\node[
    securitysub,
    below=1.0mm of ingressegress
] (l3l7policy) {
    L3--L7 Policy Enforcement
};

\begin{scope}[on background layer]
\node[
    verification,
    fit={
        (layerfourtitle)
        (ciliumpolicy)
        (podisolation)
        (networksegmentation)
        (ingressegress)
        (l3l7policy)
    },
    inner xsep=3mm,
    inner ysep=2mm
] (layerfour) {};
\end{scope}

% ==================================================================
% Layer 5: Secret Management
% ==================================================================
\node[
    font=\footnotesize\bfseries,
    text=verifygreen,
    below=3.2mm of layerfour,
    align=center
] (layerfivetitle) {
    Layer 5: Secret Management
};

\node[
    securitysub,
    below=1.0mm of layerfivetitle,
    font=\scriptsize\bfseries
] (secretmanager) {
    Secret Manager
};

\node[
    securitysub,
    below=1.0mm of secretmanager
] (secretstorage) {
    Centralized Secret Storage
};

\node[
    securitysub,
    below=1.0mm of secretstorage
] (iamaccess) {
    IAM-Controlled Access
};

\node[
    securitysub,
    below=1.0mm of iamaccess
] (auditlogs) {
    Cloud Audit Logs
};

\node[
    securitysub,
    below=1.0mm of auditlogs
] (secretversioning) {
    Secret Versioning
};

\node[
    securitysub,
    below=1.0mm of secretversioning
] (rotationworkflow) {
    Rotation Scheduling / Workflows
};

\begin{scope}[on background layer]
\node[
    verification,
    fit={
        (layerfivetitle)
        (secretmanager)
        (secretstorage)
        (iamaccess)
        (auditlogs)
        (secretversioning)
        (rotationworkflow)
    },
    inner xsep=3mm,
    inner ysep=2mm
] (layerfive) {};
\end{scope}

\end{tikzpicture}

\vspace{-2mm}
\caption{Multi-layer zero-trust security controls spanning cloud identity,
workload identity, authorization, network isolation, and secret management.}
\label{fig:zero-trust-security-stack}
\vspace{-3mm}

\end{figure*}
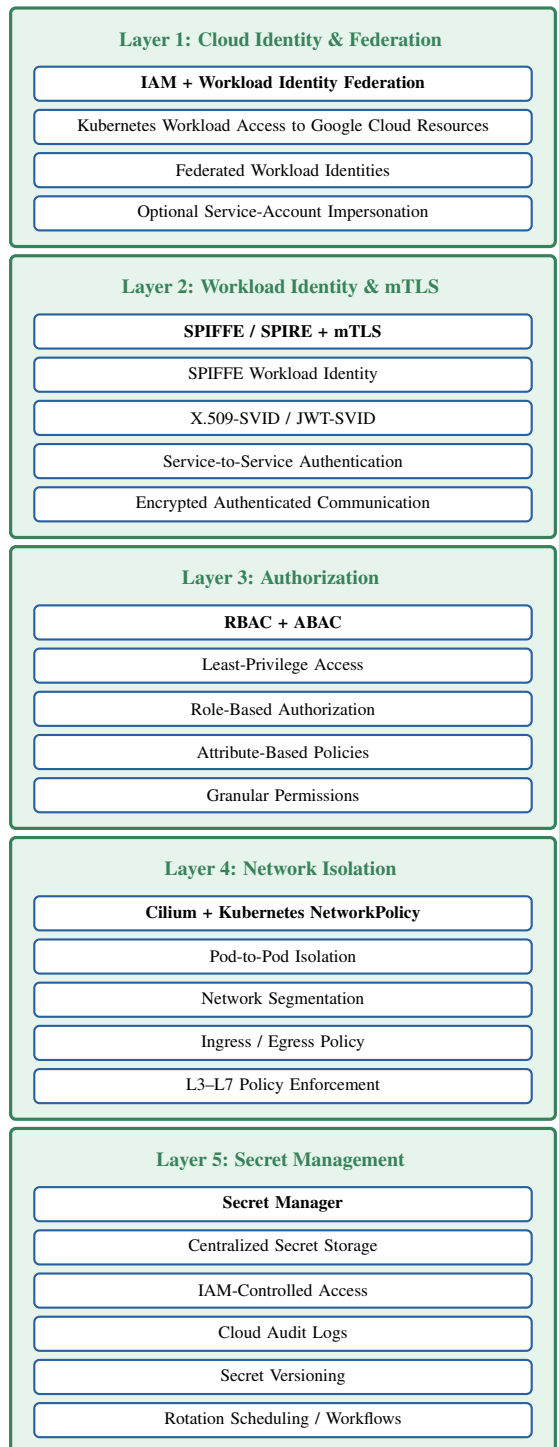

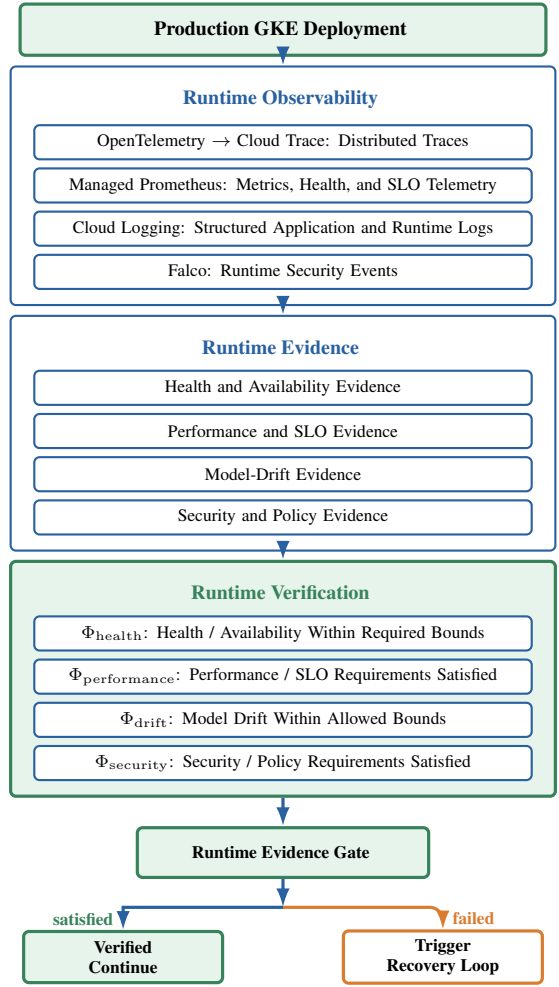
\begin{figure*}[t]
\centering

\begin{tikzpicture}[
    scale=0.94,
    transform shape,
    compactsub/.style={
        process,
        minimum width=7.0cm,
        text width=6.55cm,
        minimum height=0.48cm,
        inner xsep=3pt,
        inner ysep=1pt,
        font=\scriptsize,
        align=center
    },
    outcomeverified/.style={
        verification,
        minimum width=2.8cm,
        minimum height=0.72cm,
        font=\scriptsize\bfseries,
        align=center
    },
    outcomeloop/.style={
        rectangle,
        rounded corners=2pt,
        draw=looporange,
        very thick,
        fill=white,
        minimum width=2.8cm,
        minimum height=0.72cm,
        font=\scriptsize\bfseries,
        align=center
    }
]

% ==================================================================
% Production GKE Deployment
% ==================================================================
\node[
    verification,
    minimum width=7.4cm,
    minimum height=0.72cm,
    font=\footnotesize\bfseries
] (production) {
    Production GKE Deployment
};

% ==================================================================
% Runtime Observability
% ==================================================================
\node[
    font=\footnotesize\bfseries,
    text=graphblue,
    below=3.5mm of production,
    align=center
] (obstitle) {
    Runtime Observability
};

\node[
    compactsub,
    below=1.0mm of obstitle
] (tracing) {
    OpenTelemetry $\rightarrow$ Cloud Trace:
    Distributed Traces
};

\node[
    compactsub,
    below=1.0mm of tracing
] (metrics) {
    Managed Prometheus:
    Metrics, Health, and SLO Telemetry
};

\node[
    compactsub,
    below=1.0mm of metrics
] (logs) {
    Cloud Logging:
    Structured Application and Runtime Logs
};

\node[
    compactsub,
    below=1.0mm of logs
] (falco) {
    Falco:
    Runtime Security Events
};

\begin{scope}[on background layer]
\node[
    process,
    fit={
        (obstitle)
        (tracing)
        (metrics)
        (logs)
        (falco)
    },
    inner xsep=3mm,
    inner ysep=2mm
] (observability) {};
\end{scope}

% ==================================================================
% Runtime Evidence
% ==================================================================
\node[
    font=\footnotesize\bfseries,
    text=graphblue,
    below=3.5mm of observability,
    align=center
] (evidencetitle) {
    Runtime Evidence
};

\node[
    compactsub,
    below=1.0mm of evidencetitle
] (healthevidence) {
    Health and Availability Evidence
};

\node[
    compactsub,
    below=1.0mm of healthevidence
] (performanceevidence) {
    Performance and SLO Evidence
};

\node[
    compactsub,
    below=1.0mm of performanceevidence
] (driftevidence) {
    Model-Drift Evidence
};

\node[
    compactsub,
    below=1.0mm of driftevidence
] (securityevidence) {
    Security and Policy Evidence
};

\begin{scope}[on background layer]
\node[
    process,
    fit={
        (evidencetitle)
        (healthevidence)
        (performanceevidence)
        (driftevidence)
        (securityevidence)
    },
    inner xsep=3mm,
    inner ysep=2mm
] (runtimeevidence) {};
\end{scope}

% ==================================================================
% Runtime Verification
% ==================================================================
\node[
    font=\footnotesize\bfseries,
    text=verifygreen,
    below=3.5mm of runtimeevidence,
    align=center
] (verificationtitle) {
    Runtime Verification
};

\node[
    compactsub,
    below=1.0mm of verificationtitle
] (healthpredicate) {
    $\Phi_{\mathrm{health}}$:
    Health / Availability Within Required Bounds
};

\node[
    compactsub,
    below=1.0mm of healthpredicate
] (performancepredicate) {
    $\Phi_{\mathrm{performance}}$:
    Performance / SLO Requirements Satisfied
};

\node[
    compactsub,
    below=1.0mm of performancepredicate
] (driftpredicate) {
    $\Phi_{\mathrm{drift}}$:
    Model Drift Within Allowed Bounds
};

\node[
    compactsub,
    below=1.0mm of driftpredicate
] (securitypredicate) {
    $\Phi_{\mathrm{security}}$:
    Security / Policy Requirements Satisfied
};

\begin{scope}[on background layer]
\node[
    verification,
    fit={
        (verificationtitle)
        (healthpredicate)
        (performancepredicate)
        (driftpredicate)
        (securitypredicate)
    },
    inner xsep=3mm,
    inner ysep=2mm
] (runtimeverification) {};
\end{scope}

% ==================================================================
% Runtime Evidence Gate
% ==================================================================
\node[
    verification,
    below=4mm of runtimeverification,
    minimum width=4.1cm,
    minimum height=0.70cm,
    font=\scriptsize\bfseries
] (evidencegate) {
    Runtime Evidence Gate
};

% ==================================================================
% Final decision branch
% ==================================================================
\coordinate (decision)
    at ($(evidencegate.south)+(0,-4mm)$);

\coordinate (verifiedpos)
    at ($(decision)+(-2.25cm,-7mm)$);

\coordinate (recoverypos)
    at ($(decision)+(2.25cm,-7mm)$);

\node[
    outcomeverified
] (verifiedoutcome) at (verifiedpos) {
    Verified\\
    Continue
};

\node[
    outcomeloop
] (recoveryoutcome) at (recoverypos) {
    Trigger\\
    Recovery Loop
};

% ==================================================================
% Main vertical flow
%
% All primary downward arrows lie on the same centerline.
% ==================================================================
\draw[mainarrow]
    (production.south)
    --
    (observability.north);

\draw[mainarrow]
    (observability.south)
    --
    (runtimeevidence.north);

\draw[mainarrow]
    (runtimeevidence.south)
    --
    (runtimeverification.north);

\draw[mainarrow]
    (runtimeverification.south)
    --
    (evidencegate.north);

\draw[mainarrow]
    (evidencegate.south)
    --
    (decision);

% ==================================================================
% Evidence-gated outcomes
% ==================================================================
\draw[mainarrow]
    (decision)
    -|
    node[
        near end,
        left,
        font=\scriptsize\bfseries,
        text=verifygreen
    ] {satisfied}
    (verifiedoutcome.north);

\draw[looparrow]
    (decision)
    -|
    node[
        near end,
        right,
        font=\scriptsize\bfseries,
        text=looporange
    ] {failed}
    (recoveryoutcome.north);

\end{tikzpicture}

\vspace{-2mm}
\caption{Runtime observability, evidence derivation, and evidence-gated
verification for continued operation or bounded recovery.}
\label{fig:runtime-observability-evidence}
\vspace{-3mm}

\end{figure*}

Figures~\ref{fig:execution-agent-gke-agent-sandbox}--\ref{fig:runtime-observability-evidence}
detail the mechanisms implementing the lifecycle. Figure~\ref{fig:execution-agent-gke-agent-sandbox} separates regular GKE agent workloads from isolated artifact execution: the Execution Agent invokes the Execution MCP, generated artifacts execute in the gVisor-isolated GKE Sandbox,
and execution evidence is evaluated by the Verification Agent. Figure~\ref{fig:artifact-supply-chain-flow} shows the verified-artifact path through Cloud Build, source and container scanning, SBOM generation, signing, provenance generation, and Artifact Registry. Figure~\ref{fig:gitops-policy-promotion} details Argo CD reconciliation, OPA/Gatekeeper policy enforcement, Binary Authorization, staging deployment, live-cloud verification, and evidence-gated production promotion.
Figure~\ref{fig:zero-trust-security-stack} summarizes cloud and workload identity, authentication, authorization, network isolation, and secret management. Figure~\ref{fig:runtime-observability-evidence} shows the runtime path from telemetry collection and evidence derivation to predicate evaluation and the runtime evidence gate, which either permits continued operation or
activates bounded recovery.
%%%%%%%%%%%%%%%%%%%%%%%%%%%%%%%%%%%%%%%%%%%%%%%%%%%%%%%

%%%%%%%%%%%%%%%%%%%%%%%%%%%%%%%%%%%%%%%%%%%%%%%%%%%%%%%%%%%%%%%%%%%%%%%%%%%%%%%
% Execution Agent vs. GKE Sandbox on GKE Autopilot
%%%%%%%%%%%%%%%%%%%%%%%%%%%%%%%%%%%%%%%%%%%%%%%%%%%%%%%%%%%%%%%%%%%%%%%%%%%%%%%

\begin{figure*}[t]
\centering

\begin{tikzpicture}[
    scale=0.75,
    transform shape,
    agentnode/.style={
        process,
        minimum width=2.45cm,
        text width=2.10cm,
        minimum height=0.88cm,
        inner sep=2pt,
        font=\scriptsize,
        align=center
    },
    execagent/.style={
        graphcontrol,
        minimum width=2.55cm,
        text width=2.20cm,
        minimum height=0.92cm,
        inner sep=2pt,
        font=\scriptsize,
        align=center
    },
    verifyagent/.style={
        verification,
        minimum width=2.55cm,
        text width=2.20cm,
        minimum height=0.92cm,
        inner sep=2pt,
        font=\scriptsize,
        align=center
    },
    artifactnode/.style={
        rectangle,
        rounded corners=2pt,
        draw=black!55,
        thick,
        fill=lightgray,
        minimum width=2.45cm,
        text width=2.10cm,
        minimum height=0.82cm,
        inner sep=2pt,
        font=\scriptsize,
        align=center
    },
    protocolnode/.style={
        graphcontrol,
        minimum width=4.60cm,
        text width=4.15cm,
        minimum height=0.88cm,
        inner sep=2pt,
        font=\scriptsize,
        align=center
    },
    sandboxnode/.style={
        rectangle,
        rounded corners=3pt,
        draw=looporange,
        very thick,
        fill=white,
        minimum width=4.70cm,
        text width=4.25cm,
        minimum height=0.88cm,
        inner sep=2pt,
        font=\scriptsize,
        align=center
    },
    evidencebox/.style={
        verification,
        minimum width=3.70cm,
        text width=3.25cm,
        minimum height=0.82cm,
        inner sep=2pt,
        font=\scriptsize,
        align=center
    }
]

% =============================================================================
% GRAPH ORCHESTRATOR
% =============================================================================

\node[
    graphcontrol,
    minimum width=4.8cm,
    text width=4.35cm,
    minimum height=0.92cm,
    inner sep=2pt,
    font=\scriptsize,
    align=center
] (orchestrator) at (0,0) {
    \textbf{Graph Orchestrator}\\[-0.3mm]
    State, evidence gates, retries, recovery, termination
};

% =============================================================================
% PRIMARY AGENT PATH
% =============================================================================

\node[
    agentnode
] (generation) at (-6.2,-2.0) {
    \textbf{Repository\\Generation Agent}
};

\node[
    artifactnode
] (repository) at (-3.5,-2.0) {
    \textbf{Generated\\Repository}
};

\node[
    agentnode
] (review) at (-0.8,-2.0) {
    \textbf{Repository\\Review Agent}
};

\node[
    execagent
] (execution) at (2.0,-2.0) {
    \textbf{Execution Agent}\\[-0.3mm]
    Controls execution
};

\node[
    verifyagent
] (verificationagent) at (5.0,-2.0) {
    \textbf{Verification Agent}\\[-0.3mm]
    Evaluates evidence
};

% =============================================================================
% CORRECTION AGENTS
% =============================================================================

\node[
    agentnode
] (reflection) at (-4.8,-4.0) {
    \textbf{Reflection Agent}\\[-0.3mm]
    Diagnosis
};

\node[
    agentnode
] (repair) at (-1.8,-4.0) {
    \textbf{Repair Agent}\\[-0.3mm]
    Artifact correction
};

% =============================================================================
% RELEASE AND MONITORING AGENTS
% =============================================================================

\node[
    agentnode
] (release) at (5.0,-4.0) {
    \textbf{Release Agent}
};

\node[
    agentnode
] (monitoring) at (7.8,-4.0) {
    \textbf{Monitoring Agent}\\[-0.3mm]
    Runtime evidence
};

% =============================================================================
% GRAPH ORCHESTRATOR -> PRIMARY PATH
%
% Dedicated left-side route prevents the arrow from passing through
% the repository or agent boxes.
% =============================================================================

\coordinate (orchestratorlane)
    at (-7.7,0);

\coordinate (generationentry)
    at (-7.7,-2.0);

\draw[mainarrow]
    (orchestrator.west)
    -- (orchestratorlane)
    |- (generation.west);

% =============================================================================
% PRIMARY REPOSITORY FLOW
% =============================================================================

\draw[mainarrow]
    (generation.east)
    -- (repository.west);

\draw[mainarrow]
    (repository.east)
    -- (review.west);

\draw[mainarrow]
    (review.east)
    -- (execution.west);

% =============================================================================
% REFLECTION -> REPAIR
% =============================================================================

\draw[looparrow]
    (reflection.east)
    -- node[
        looplabel,
        above
    ] {diagnosis}
    (repair.west);

% =============================================================================
% REPAIR -> REVIEW
%
% Dedicated vertical re-entry corridor between the repository and review
% nodes. No line passes through either box.
% =============================================================================

\coordinate (repairup)
    at (-1.8,-3.15);

\coordinate (reviewdown)
    at (-0.8,-3.15);

\draw[looparrow]
    (repair.north)
    -- (repairup)
    -- node[
        looplabel,
        above
    ] {correct}
    (reviewdown)
    -- (review.south);

% =============================================================================
% VERIFIED -> RELEASE -> MONITORING
% =============================================================================

\draw[
    ->,
    very thick,
    draw=verifygreen
]
    (verificationagent.south)
    -- node[
        right,
        fill=white,
        inner sep=1.5pt,
        font=\scriptsize,
        text=verifygreen
    ] {verified}
    (release.north);

\draw[mainarrow]
    (release.east)
    -- (monitoring.west);

% =============================================================================
% AGENTIC CONTROL / APPLICATION LAYER BOUNDARY
% =============================================================================

\begin{scope}[on background layer]

\node[
    draw=graphblue,
    dashed,
    very thick,
    rounded corners=5pt,
    inner xsep=4mm,
    inner ysep=4mm,
    fit={
        (orchestrator)
        (generation)
        (repository)
        (review)
        (execution)
        (verificationagent)
        (reflection)
        (repair)
        (release)
        (monitoring)
    }
] (agentlayer) {};

\end{scope}

\node[
    anchor=west,
    fill=white,
    inner xsep=2pt,
    inner ysep=1pt,
    text=graphblue,
    font=\scriptsize\bfseries
] at ($(agentlayer.north west)+(3mm,0)$) {
    Agentic Control / Application Layer
};

\node[
    anchor=east,
    fill=white,
    inner xsep=2pt,
    inner ysep=1pt,
    text=graphblue,
    font=\scriptsize
] at ($(agentlayer.north east)+(-3mm,0)$) {
    Regular GKE Workloads
};

% =============================================================================
% EXECUTION MCP
%
% A clear vertical corridor is intentionally retained directly below
% the Execution Agent.
% =============================================================================

\node[
    protocolnode
] (executionmcp) at (2.0,-6.0) {
    \textbf{Execution MCP}\\[-0.3mm]
    Policy-scoped access to controlled execution
};

\draw[mainarrow]
    (execution.south)
    -- (executionmcp.north);

% Separate label placed beside the arrow rather than on the arrow.
\node[
    anchor=west,
    fill=white,
    inner sep=1.5pt,
    text=graphblue,
    font=\scriptsize
] at (2.25,-5.0) {
    execution request
};

% =============================================================================
% GKE SANDBOX
% =============================================================================

\node[
    sandboxnode
] (gkesandbox) at (2.0,-7.8) {
    \textbf{GKE Sandbox with gVisor}\\[-0.3mm]
    Isolated execution environment
};

\node[
    sandboxnode
] (artifactexecution) at (2.0,-9.15) {
    \textbf{Controlled Artifact Execution}\\[-0.3mm]
    Generated artifacts execute in isolation
};

% =============================================================================
% GKE SANDBOX BOUNDARY
% =============================================================================

\begin{scope}[on background layer]

\node[
    draw=looporange,
    dashed,
    very thick,
    rounded corners=5pt,
    inner xsep=4mm,
    inner ysep=4mm,
    fit={
        (gkesandbox)
        (artifactexecution)
    }
] (sandboxboundary) {};

\end{scope}

\node[
    anchor=west,
    fill=white,
    inner xsep=2pt,
    inner ysep=1pt,
    text=looporange,
    font=\scriptsize\bfseries
] at ($(sandboxboundary.north west)+(3mm,0)$) {
    Isolated Execution
};

% =============================================================================
% EXECUTION MCP -> GKE SANDBOX
% =============================================================================

\draw[mainarrow]
    (executionmcp.south)
    -- (gkesandbox.north);

% Label is offset from the arrow.
\node[
    anchor=west,
    fill=white,
    inner sep=1.5pt,
    text=graphblue,
    font=\scriptsize
] at (2.25,-6.92) {
    controlled access
};

% =============================================================================
% SANDBOX EXECUTION
% =============================================================================

\draw[mainarrow]
    (gkesandbox.south)
    -- (artifactexecution.north);

% =============================================================================
% EXECUTION EVIDENCE
% =============================================================================

\node[
    evidencebox
] (evidence) at (2.0,-10.75) {
    \textbf{Execution Evidence}
};

\draw[mainarrow]
    (artifactexecution.south)
    -- (evidence.north);

% =============================================================================
% EXECUTION EVIDENCE -> VERIFICATION AGENT
%
% This route remains completely outside the right edge of all nodes.
% =============================================================================

\coordinate (evidencelane)
    at (9.5,-10.75);

\coordinate (verificationlane)
    at (9.5,-2.0);

\draw[
    ->,
    very thick,
    draw=verifygreen,
    rounded corners=5pt
]
    (evidence.east)
    -- (evidencelane)
    -- (verificationlane)
    -- (verificationagent.east);

\node[
    anchor=west,
    fill=white,
    inner sep=1.5pt,
    text=verifygreen,
    font=\scriptsize
] at (9.55,-6.4) {
    execution evidence
};

% =============================================================================
% FAILURE / CORRECTION INDICATION
%
% Instead of routing a long arrow across the agent boxes, verification
% failure is represented by a short local marker connected to the bounded
% correction pair. This keeps the execution path visually clean.
% =============================================================================

\coordinate (failurestart)
    at ($(verificationagent.north)+(0,5mm)$);

\coordinate (failureleft)
    at (-4.8,-0.9);

\coordinate (reflectionentry)
    at (-4.8,-3.35);

\draw[looparrow]
    (verificationagent.north)
    -- (failurestart)
    -- node[
        looplabel,
        above,
        pos=0.42
    ] {predicate failed}
    (failureleft)
    -- (reflectionentry)
    -- (reflection.north);

% =============================================================================
% RUNTIME FEEDBACK
%
% Routed entirely around the right and bottom edges of the agent layer.
% It does not pass through the Execution Agent or Execution MCP corridor.
% =============================================================================

\coordinate (runtimeouter)
    at (9.0,-4.0);

\coordinate (runtimebottomright)
    at (9.0,-5.05);

\coordinate (runtimebottomleft)
    at (-4.8,-5.05);

\draw[looparrow]
    (monitoring.east)
    -- (runtimeouter)
    -- (runtimebottomright)
    -- node[
        looplabel,
        below,
        pos=0.52
    ] {runtime trigger}
    (runtimebottomleft)
    -- (reflection.south);

% =============================================================================
% OUTER GKE AUTOPILOT CLUSTER
% =============================================================================

\begin{scope}[on background layer]

\node[
    draw=black!65,
    very thick,
    rounded corners=6pt,
    inner xsep=6mm,
    inner ysep=6mm,
    fit={
        (agentlayer)
        (executionmcp)
        (sandboxboundary)
        (evidence)
        (evidencelane)
        (orchestratorlane)
        (runtimeouter)
        (runtimebottomleft)
    }
] (autopilot) {};

\end{scope}

\node[
    anchor=west,
    fill=white,
    inner xsep=3pt,
    inner ysep=1pt,
    font=\small\bfseries
] at ($(autopilot.north west)+(4mm,0)$) {
    GKE Autopilot Cluster
};

\end{tikzpicture}

\vspace{-2mm}

\caption{The figure shows the execution-agent and sandbox separation on GKE Autopilot. The Graph
Orchestrator and specialized agents run as regular GKE workloads. The Execution Agent requests controlled execution through the Execution MCP, whereas the GKE Sandbox with gVisor provides the isolated environment for generated-artifact execution. Execution evidence is evaluated by the
Verification Agent; failed verification activates bounded reflection and repair, while verified artifacts proceed to release and monitoring.
}
\label{fig:execution-agent-gke-agent-sandbox}

\vspace{-3mm}
\end{figure*}

%%%%%%%%%%%%%%%%%%%%%%%%%%%%%%%%%%%%%%%%%%%%%%%%%%%%%%%%%%%%%%%%%%%%%%%%%%%%%%%

%%%%%%%%%%%%%%%%%%%%%%%%%%%%%%%%%%%%%%%%%%%%%
%%%%%%%%%%%%%%%%%%%%%%%%%%%%%%%%%%%%%%%%%%%%%

%%%%%%%%%%%%%%%%%%%%%%%%%%%%%%%%%%%%%%%%%%%%%
\begin{figure*}[t]
\centering

\begin{tikzpicture}[
    node distance=4.5mm,
    substepwide/.style={
        process,
        minimum width=6.8cm,
        text width=6.35cm,
        minimum height=0.56cm,
        inner ysep=2pt,
        font=\scriptsize,
        align=center
    },
    substepnarrow/.style={
        process,
        minimum width=3.25cm,
        text width=2.85cm,
        minimum height=0.56cm,
        inner ysep=2pt,
        font=\scriptsize,
        align=center
    }
]

% ==================================================================
% MAIN CENTERLINE
%
% All principal lifecycle stages are centered on x = 0.
% ==================================================================
\coordinate (mainaxis) at (0,0);

% ==================================================================
% Argo CD
% ==================================================================
\node[
    font=\footnotesize\bfseries,
    text=graphblue,
    align=center
] (argotitle) at (mainaxis) {
    Argo CD
};

\node[
    substepwide,
    below=2.5mm of argotitle
] (gitreposync) {
    Git Repository Synchronization
};

\node[
    substepwide,
    below=2.5mm of gitreposync
] (desiredstate) {
    Desired-State Reconciliation
};

\node[
    substepwide,
    below=2.5mm of desiredstate
] (driftdetection) {
    Drift Detection
};

\node[
    substepwide,
    below=2.5mm of driftdetection
] (automatedsync) {
    Automated Synchronization
};

\begin{scope}[on background layer]
\node[
    process,
    fit={
        (argotitle)
        (gitreposync)
        (desiredstate)
        (driftdetection)
        (automatedsync)
    },
    inner xsep=4mm,
    inner ysep=3mm
] (argocd) {};
\end{scope}

% ==================================================================
% CENTRAL POLICY FORK
%
% This coordinate lies exactly on the x = 0 centerline.
% ==================================================================
\coordinate (policyfork)
    at ($(argocd.south)+(0,-4mm)$);

% ==================================================================
% Parallel policy-control positions
%
% Both branches are placed symmetrically around x = 0.
% ==================================================================
\coordinate (policyleft)
    at ($(policyfork)+(-3.0cm,-8mm)$);

\coordinate (policyright)
    at ($(policyfork)+(3.0cm,-8mm)$);

% ==================================================================
% OPA / Gatekeeper
% ==================================================================
\node[
    font=\footnotesize\bfseries,
    text=graphblue,
    align=center
] (opatitle) at (policyleft) {
    OPA / Gatekeeper
};

\node[
    substepnarrow,
    below=2.5mm of opatitle
] (policyascode) {
    Policy as Code
};

\node[
    substepnarrow,
    below=2.5mm of policyascode
] (admissioncontrol) {
    Admission Control
};

\node[
    substepnarrow,
    below=2.5mm of admissioncontrol
] (constraintenforcement) {
    Constraint Enforcement
};

\node[
    substepnarrow,
    below=2.5mm of constraintenforcement
] (custompolicies) {
    Custom Policies
};

\begin{scope}[on background layer]
\node[
    process,
    fit={
        (opatitle)
        (policyascode)
        (admissioncontrol)
        (constraintenforcement)
        (custompolicies)
    },
    inner xsep=4mm,
    inner ysep=3mm
] (opabox) {};
\end{scope}

% ==================================================================
% Binary Authorization
% ==================================================================
\node[
    font=\footnotesize\bfseries,
    text=graphblue,
    align=center
] (bintitle) at (policyright) {
    Binary Authorization
};

\node[
    substepnarrow,
    below=2.5mm of bintitle
] (attestationpolicy) {
    Attestation Policy
};

\node[
    substepnarrow,
    below=2.5mm of attestationpolicy
] (attestationverification) {
    Attestation Verification
};

\node[
    substepnarrow,
    below=2.5mm of attestationverification
] (deploymentenforcement) {
    Deployment Enforcement
};

\begin{scope}[on background layer]
\node[
    process,
    fit={
        (bintitle)
        (attestationpolicy)
        (attestationverification)
        (deploymentenforcement)
    },
    inner xsep=4mm,
    inner ysep=3mm
] (binaryauth) {};
\end{scope}

% ==================================================================
% CENTRAL POLICY JOIN
%
% OPA/Gatekeeper is the taller branch, so its bottom defines the
% safe merge level. The x-coordinate is inherited from policyfork,
% keeping the join exactly on the main centerline.
% ==================================================================
\coordinate (policyjoinlevel)
    at ($(opabox.south)+(0,-4mm)$);

\coordinate (policyjoin)
    at (policyfork |- policyjoinlevel);

% ==================================================================
% Staging GKE Deployment
%
% stagingtitle is placed explicitly on the same x = 0 centerline.
% ==================================================================
\coordinate (stagingcenter)
    at ($(policyjoin)+(0,-10mm)$);

\node[
    font=\footnotesize\bfseries,
    text=verifygreen,
    align=center
] (stagingtitle) at (stagingcenter) {
    Staging GKE Deployment
};

\node[
    substepwide,
    below=2.5mm of stagingtitle
] (stagingservices) {
    Frontend + Backend + Inference
};

\node[
    substepwide,
    below=2.5mm of stagingservices
] (namespaceisolation) {
    Namespace Isolation
};

\node[
    substepwide,
    below=2.5mm of namespaceisolation
] (resourcelimits) {
    Resource Limits
};

\begin{scope}[on background layer]
\node[
    verification,
    fit={
        (stagingtitle)
        (stagingservices)
        (namespaceisolation)
        (resourcelimits)
    },
    inner xsep=4mm,
    inner ysep=3mm
] (staging) {};
\end{scope}

% ==================================================================
% Live-Cloud Verification
%
% Positioned directly below Staging; therefore both boxes have
% exactly the same horizontal center.
% ==================================================================
\node[
    font=\footnotesize\bfseries,
    text=verifygreen,
    below=8mm of staging,
    align=center
] (verificationtitle) {
    Live-Cloud Verification
};

\node[
    substepwide,
    below=2.5mm of verificationtitle
] (healthchecks) {
    Health Checks
};

\node[
    substepwide,
    below=2.5mm of healthchecks
] (integrationtests) {
    Integration Tests
};

\node[
    substepwide,
    below=2.5mm of integrationtests
] (performancetests) {
    Performance Tests
};

\node[
    substepwide,
    below=2.5mm of performancetests
] (securityvalidation) {
    Security Validation
};

\begin{scope}[on background layer]
\node[
    verification,
    fit={
        (verificationtitle)
        (healthchecks)
        (integrationtests)
        (performancetests)
        (securityvalidation)
    },
    inner xsep=4mm,
    inner ysep=3mm
] (liveverification) {};
\end{scope}

% ==================================================================
% Production GKE Deployment
%
% Positioned directly below Live-Cloud Verification; therefore the
% entire promotion path remains on one vertical centerline.
% ==================================================================
\node[
    font=\footnotesize\bfseries,
    text=verifygreen,
    below=10mm of liveverification,
    align=center
] (productiontitle) {
    Production GKE Deployment
};

\node[
    substepwide,
    below=2.5mm of productiontitle
] (productionservices) {
    Frontend + Backend + Inference
};

\node[
    substepwide,
    below=2.5mm of productionservices
] (productionconfig) {
    Production Configuration
};

\node[
    substepwide,
    below=2.5mm of productionconfig
] (scalingavailability) {
    Scaling and Availability Controls
};

\begin{scope}[on background layer]
\node[
    verification,
    fit={
        (productiontitle)
        (productionservices)
        (productionconfig)
        (scalingavailability)
    },
    inner xsep=4mm,
    inner ysep=3mm
] (production) {};
\end{scope}

% ==================================================================
% FLOW: Argo CD -> central policy fork
%
% Exact vertical line on x = 0.
% ==================================================================
\draw[mainarrow]
    (argocd.south)
    --
    (policyfork);

% ==================================================================
% FLOW: central fork -> parallel policy controls
%
% The branches leave the main centerline symmetrically.
% ==================================================================
\draw[mainarrow]
    (policyfork)
    -|
    (opabox.north);

\draw[mainarrow]
    (policyfork)
    -|
    (binaryauth.north);

% ==================================================================
% FLOW: policy controls -> central join
%
% These are merge lines rather than terminal transitions, so only
% the single central outgoing edge below uses an arrowhead.
% ==================================================================
\draw[
    very thick,
    draw=graphblue
]
    (opabox.south)
    |-
    (policyjoin);

\draw[
    very thick,
    draw=graphblue
]
    (binaryauth.south)
    |-
    (policyjoin);

% ==================================================================
% FLOW: central policy join -> Staging
%
% Exact vertical line on x = 0.
% ==================================================================
\draw[mainarrow]
    (policyjoin)
    --
    (staging.north);

% ==================================================================
% FLOW: Staging -> Live-Cloud Verification
%
% Exact vertical line on x = 0.
% ==================================================================
\draw[mainarrow]
    (staging.south)
    --
    (liveverification.north);

% ==================================================================
% FLOW: Live-Cloud Verification -> Production
%
% Exact vertical line on x = 0.
% ==================================================================
\draw[mainarrow]
    (liveverification.south)
    --
    node[
        right,
        font=\scriptsize\bfseries,
        text=verifygreen
    ] {verified}
    (production.north);

\end{tikzpicture}

\caption{GitOps reconciliation, policy enforcement, and evidence-gated promotion from staging to production.}
\label{fig:gitops-policy-promotion}

\end{figure*}
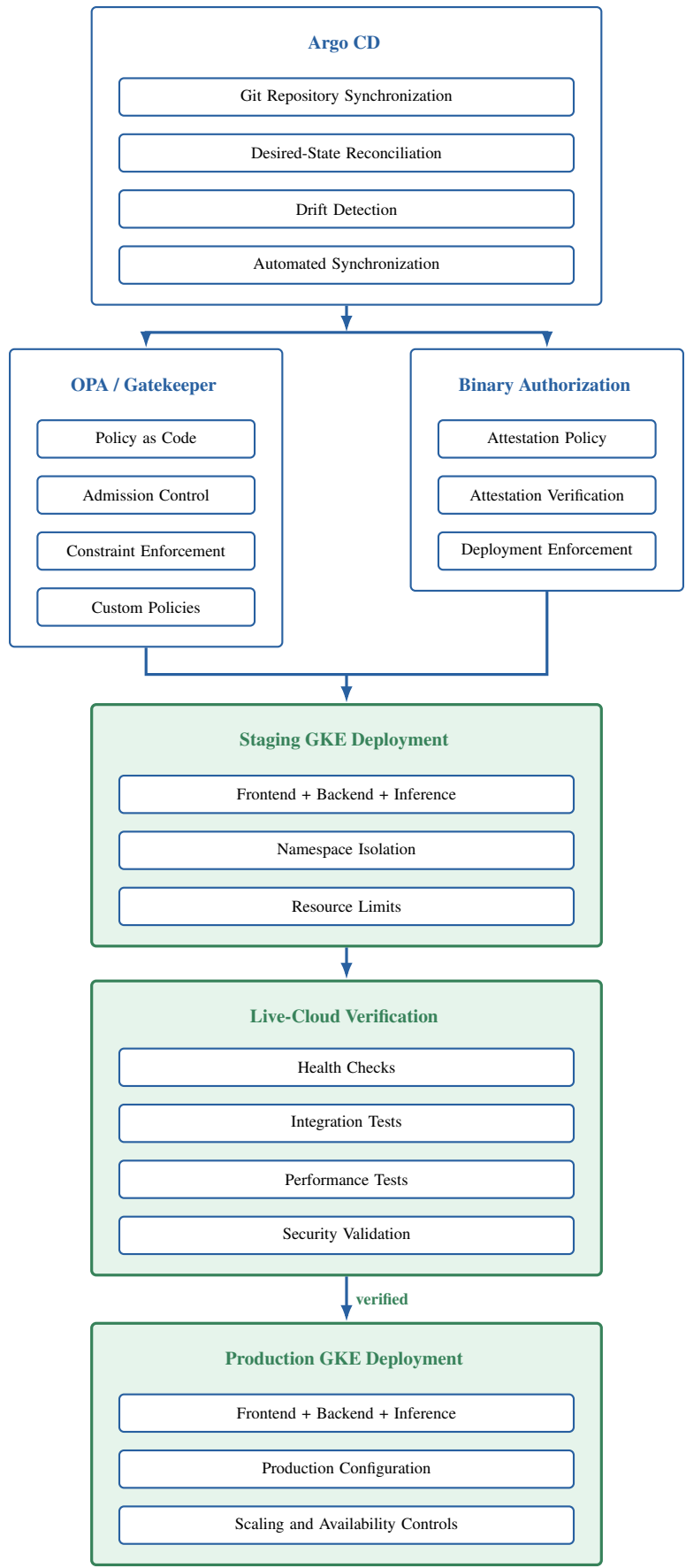
%%%%%%%%%%%%%%%%%%%%%%%%%%%%%%%%%%%%%%%%%%%%%

%%%%%%%%%%%%%%%%%%%%%%%%%%%%%%%%%%%%%%%%%%%%%
%%%%%%%%%%%%%%%%%%%%%%%%%%%%%%%%%%%%%%%%%%%%%

%%%%%%%%%%%%%%%%%%%%%%%%%%%%%%%%%%%%%%%%%%%%%
%%%%%%%%%%%%%%%%%%%%%%%%%%%%%%%%%%%%%%%%%%%%%

\end{document}